\documentclass[prb,preprint,showpacs,eqsecnum]{revtex4}
\usepackage{graphicx}
\usepackage{bm}
\begin{document}
\title{Resonant suppression of thermal stability of the nanoparticle
magnetization by a rotating magnetic field}
\author{S.~I.~Denisov, A.~Yu.~Polyakov, and T.~V.~Lyutyy}
\affiliation{Sumy State University, 2 Rimsky-Korsakov Street, 40007
Sumy, Ukraine}

%\date{submitted to Physical Review B: \today}

\begin{abstract}
We study the thermal stability of the periodic (P) and quasi-periodic
(Q) precessional modes of the nanoparticle magnetic moment induced by
a rotating magnetic field. An analytical method for determining the
lifetime of the P mode in the case of high anisotropy barrier and
small amplitudes of the rotating field is developed within the
Fokker-Planck formalism. In general case, the thermal stability of
both P and Q modes is investigated by numerical simulation of the
stochastic Landau-Lifshitz equation. We show analytically and
numerically that the lifetime is a nonmonotonic function of the
rotating field frequency which, depending on the direction of field
rotation, has either a pronounced maximum or a deep minimum near the
Larmor frequency.
\end{abstract}
 \maketitle

\section{INTRODUCTION}
\label{Intr}

Magnetic nanoparticles are of great interest because of their
nanoscale physical properties and many current and potential
applications. These applications range from high-density storage
media\cite{MTMA, Ross} and spintronic devices\cite{ZFS, WABD} to
biomedical applications like drug delivery, cell separation, cancer
treatment and many others (for a review, see Refs.~\onlinecite{LFPR,
Ferr, PCJD}). Since the physical properties of nanoparticles play a
decisive role in all these applications, their study is of
fundamental importance. In particular, for high-density storage
media, e.g., bit-patterned media\cite{Ross, Richt} where each
nanoparticle is a carrier of information, the thermal stability of a
given direction (or magnitude) of the nanoparticle magnetic moment is
one of the most important problems. The reason is that under thermal
fluctuations the magnetic moment can be randomly switched to a new
state leading to the loss of information.

In the case of ferromagnetic nanoparticles, the fluctuation dynamics
of the nanoparticle magnetic moment can be described by the
stochastic Landau-Lifshitz equation. If the noise term in this
equation is approximated by the Gaussian white noise, then the
probability density of the magnetic moment satisfies the
Fokker-Planck equation whose properties are well known.\cite{Ris}
This approach, introduced by Brown\cite{Brown} almost five decades
ago, has become an important tool in the study of stochastic magnetic
dynamics (see Ref.~\onlinecite{CKW} and references therein). To
characterize the thermal stability in the case of uniaxial
nanoparticles, it is often enough to determine the lifetimes of the
nanoparticle magnetic moment in the ``up" and ``down" states. In the
above approach, the lifetime in a given state is usually interpreted
as the relaxation time. However, from a theoretical point of view,
the lifetime is reasonable to associate with the mean first-passage
time (MFPT), i.e., average time that a random process dwells in a
prescribed state. An additional advantageous feature of this
definition of the lifetime is that the MFPT method is mathematically
well developed.\cite{HTB, Gar, Red} This approach was first applied
to study the magnetic relaxation in systems of
noninteracting\cite{DY} and dipolar interacting\cite{DLT}
nanoparticles subjected to a constant magnetic field.

In general, the lifetime depends on both intrinsic properties of
nanoparticles and external magnetic fields. The case when the
external fields contain a rotating magnetic field applied
perpendicular to the easy axes of nanoparticles has a particular
interest. On the one hand, this is because the rotating field plays a
key role in the microwave-assisted switching\cite{TWM, SW1, SW2, PH,
WB, ZZT} and, on the other hand, because the corresponding dynamical
equations (without accounting the thermal fluctuations) can often be
solved analytically.\cite{ BSM,BMS1, DLHT, BMSAB, CSS} Specifically,
it has been shown\cite{ BSM} (see also Ref.~\onlinecite{BMS}) that
the rotating field can induce two types of the stable precessional
modes of the magnetic moment, namely, the periodic (P) and
quasi-periodic (Q) modes. Under certain conditions,\cite{DLBH, LPRB}
one mode can exist in the up state of the magnetic moment and the
other in the down state. The thermal fluctuations make the random
transitions between these modes possible, and the problem of the
lifetime of the P and Q modes appears. Some aspects of this problem
have already been considered previously in the context of magnetic
relaxation and induced magnetization.\cite{DLH, DSTH1, DSTH2} But the
dependence of the lifetime on the parameters of the rotating field
has not been studied systematically. At the same time, the effect of
strong dependence of the lifetime on the rotating field frequency,
which is expected to exist in the vicinity of the Larmor frequency,
could be important for applications. Therefore, in this paper we
present a detailed analytical and numerical analysis of the above
mentioned problem.

The paper is organized as follows. In Sec.~\ref{Life}, we describe
the model and define the lifetime of both the P and Q modes. Here we
also derive the boundary conditions and transformation properties of
the lifetime. In Sec.~\ref{AnRes}, we develop an analytical method
for calculating the frequency dependence of the lifetime of the P
mode in the case of high anisotropy barrier. Our numerical results
obtained by the simulation of the deterministic and stochastic
Landau-Lifshitz equations are presented in Sec.~\ref{NumRes}.
Specifically, the features of the P and Q modes at zero temperature
are studied in Sec.~\ref{Prec} and the effects of thermal
fluctuations are considered in Sec.~\ref{Simulat}. Finally, in
Sec.~\ref{Con} we summarize our findings.

\section{LIFETIMES OF THE PRECESSIONAL MODES}
\label{Life}
\subsection{Basic equations of the model}
\label{Model}

To study the influence of the rotating magnetic field on the thermal
stability of the nanoparticle magnetization, we use a minimal model
with coherent spin dynamics. Within this model, which is applicable
to particles whose exchange energy is comparatively large, the
magnetic state of each particle is completely characterized by the
magnetic moment $\mathbf{m} = \mathbf{m}(t)$ of a fixed magnitude $m
= |\mathbf{m}|$. Due to its interaction with a heat bath,
$\mathbf{m}(t)$ is a vector random process which can be described by
the stochastic Landau-Lifshitz equation\cite{KH}
\begin{equation}
    \frac{d}{dt}\mathbf{m} = -\gamma \mathbf{m} \times
    (\mathbf{H} + \mathbf{n}) - \frac{\lambda\gamma}{m}\,
    \mathbf{m} \times (\mathbf{m} \times \mathbf{H}),
    \label{L-L}
\end{equation}
where $\gamma(>0)$ is the gyromagnetic ratio, $\lambda(>0)$ is the
damping parameter, the cross denotes the vector product, and
$\mathbf{H} = \mathbf{H} (t)$ and $\mathbf{n} = \mathbf{n}(t)$ are
the effective magnetic fields. The first field is given by
$\mathbf{H} = -\partial W/ \partial \mathbf{m}$, where $W$ is the
magnetic energy of the particle which, in the case under
consideration, contains only the uniaxial anisotropy energy
$(1/2)H_{a}m (1 - m_{z}^{2}/m^{2})$ and the Zeeman energy
$-\mathbf{m}\cdot \mathbf{h}(t)$, i.e., $\mathbf{H} =
H_{a}(m_{z}/m)\mathbf{e}_{z} + \mathbf{h}(t)$. Here, the $z$-axis of
a Cartesian coordinate system $xyz$ with unit vectors
$\mathbf{e}_{x}$, $\mathbf{e}_{y}$ and $\mathbf{e}_{z}$ is chosen to
be parallel to the easy axis of magnetization, $H_{a}(>0)$ is the
anisotropy field, $m_{z} = \mathbf{m}\cdot \mathbf{e}_{z}$, the dot
denotes the scalar product, and $\mathbf{h}(t)$ is the rotating
magnetic field. We assume that $\mathbf{h}(t)$ is applied
perpendicular to the $z$-axis, so that
\begin{equation}
    \mathbf{h}(t) = h\cos(\omega t)\mathbf{e}_{x} +
    \rho h\sin(\omega t)\mathbf{e}_{y},
    \label{def h}
\end{equation}
where $h = |\mathbf{h}(t)|$ is the field amplitude, $\omega$ is the
angular rotation frequency, and $\rho=-1$ (for clockwise rotation) or
$+1$ (for counterclockwise rotation). It is this field that induces
the precessional modes of $\mathbf{m}(t)$.

The effective field $\mathbf{n}(t)$ accounts for the thermal
fluctuations. It is assumed that the Cartesian components $n_{\alpha}
(t)$ $(\alpha = x,y,z)$ of $\mathbf{n}(t)$ are independent Gaussian
white noises with zero means and correlation functions $\langle
n_{\alpha}(t_{1}) n_{\alpha}(t_{2}) \rangle = 2 \Delta \delta(t_{2} -
t_{1})$. Here, the angular brackets denote averaging over all
realizations of $\mathbf{n}(t)$, $\Delta = \lambda k_{B}T/\gamma m$
is the noise intensity, $k_{B}$ is the Boltzmann constant, $T$ is the
absolute temperature, and $\delta(t)$ is the Dirac $\delta$ function.
In accordance with this definition of $\mathbf{n}(t)$, the random
process $\mathbf{m}(t)$ is Markovian and can be described within the
Fokker-Planck formalism.

Because Eq.~(\ref{L-L}) preserves the length of the magnetic moment
$\mathbf{m}(t)$, it is convenient to write the Fokker-Planck equation
that corresponds to Eq.~(\ref{L-L}) in spherical coordinates.
Introducing the polar and azimuthal angles $\theta(t)$ and
$\varphi(t)$ of $\mathbf{m}(t)$, the forward and backward
Fokker-Planck equations for the conditional probability density $P =
P(\theta,\psi, \tilde{t} |\theta',\psi',\tilde{t}')$ ($\tilde{t} \geq
\tilde{t}'$) can be written as\cite{DSTH1,DSTH2}
\begin{eqnarray}
    &\displaystyle \frac{\partial^{2}P}{\partial\theta^{2}} +
    \frac{1} {\sin^{2}\theta}\frac{\partial^{2}P}{\partial\psi^{2}}
    - \frac{\partial}{\partial\theta}\bigg(\cot\theta +
    \frac{2a}{\lambda} u(\theta,\psi) \bigg)P&
    \nonumber\\[6pt]
    &\displaystyle -\frac{2a}{\lambda} \frac{\partial}
    {\partial\psi} [v(\theta,\psi) - \rho \tilde{\omega}]P =
    \frac{2a}{\lambda} \frac{\partial P}{\partial \tilde{t}}&
    \label{F-Pfw}
\end{eqnarray}
and
\begin{eqnarray}
    &\displaystyle \frac{\partial^{2}P}{\partial\theta'^{2}} +
    \frac{1}
    {\sin^{2}\theta'}\frac{\partial^{2}P}{\partial\psi'^{2}}
    + \bigg(\cot\theta' + \frac{2a}{\lambda} u(\theta',\psi')
    \bigg) \frac{\partial P}{\partial\theta'}&
    \nonumber\\[6pt]
    &\displaystyle +\frac{2a}{\lambda} [v(\theta',\psi') -
    \rho \tilde{\omega}]\frac{\partial P}{\partial\psi'} =
    -\frac{2a}{\lambda} \frac{\partial P}{\partial
    \tilde{t}'},&
    \label{F-Pbw}
\end{eqnarray}
respectively. Here, $\tilde{t} = \omega_{r}t$ is the dimensionless
time, $\omega_{r} = \gamma H_{a}$ is the Larmor frequency,
$\tilde{\omega} = \omega/\omega_{r}$ is the dimensionless frequency
of the rotating field, and $a=mH_{a}/2k_{B}T$ is a dimensionless
parameter that characterizes the anisotropy barrier height in the
units of the thermal energy $k_{B}T$. Finally, the variables $\theta$
and $\psi$ are associated with $\theta(\tilde{t})$ and
$\psi(\tilde{t}) = \varphi (\tilde{t}) - \rho \tilde{\omega}
\tilde{t}$, respectively, and the functions $u(\theta,\psi)$ and
$v(\theta,\psi)$ are expressed through the dimensionless magnetic
energy
\begin{equation}
    \tilde{W} = \frac{W}{mH_{a}} = \frac{1}{2}\sin^2{\theta}
    - \tilde{h}\sin{\theta} \cos{\psi}
    \label{defW}
\end{equation}
($\tilde{h} = h/H_{a}$) as follows:
\begin{eqnarray}
    u(\theta,\psi) \!&=&\! -\frac{1}{\sin \theta} \bigg(
    \lambda \sin \theta \frac{\partial}{\partial \theta} +
    \frac{\partial}{\partial \psi} \bigg) \tilde{W}
    \nonumber\\
    \!&=&\! -\lambda\sin\theta\cos\theta - \tilde{h}\sin\psi
    + \lambda\tilde{h}\cos\theta\cos\psi,
    \nonumber\\[6pt]
    v(\theta,\psi) \!&=&\! \frac{1}{\sin^{2} \theta} \bigg(
    \sin \theta \frac{\partial}{\partial \theta} -
    \lambda \frac{\partial}{\partial \psi} \bigg) \tilde{W}
    \nonumber\\
    \!&=&\! \cos\theta - \tilde{h}\cot\theta\cos\psi -
    \lambda\tilde{h}\frac{\sin\psi}{\sin\theta}.
    \label{def uv}
\end{eqnarray}
It is assumed that the probability density $P$ satisfies the initial
condition $P|_{\tilde{t} = \tilde{t}'} = \delta(\theta - \theta')
\delta(\psi - \psi')$. Moreover, if the absorbing boundary conditions
are not imposed, then $P$ is properly normalized: $\int_{0}^{2\pi}
d\psi \int_{0}^{\pi} d\theta P =1$.

The system of two stochastic differential equations
\begin{eqnarray}
    &\displaystyle \frac{d}{d\tilde{t}}\theta(\tilde{t}) =
    u\big(\theta(\tilde{t}), \psi(\tilde{t})\big) +
    \frac{\lambda}{2a} \cot \theta(\tilde{t}) +
    \sqrt{\frac{\lambda}{a}}\, \eta_{1}(\tilde{t}),&
    \nonumber \\[4pt]
    &\displaystyle \frac{d}{d\tilde{t}}\psi(\tilde{t}) =
    v\big(\theta(\tilde{t}), \psi(\tilde{t})\big) -
    \rho \tilde{\omega} + \sqrt{\frac{\lambda}
    {a}}\,\frac{1}{\sin \theta(\tilde{t})} \,
    \eta_{2}(\tilde{t}),&\qquad
    \label{L-L2}
\end{eqnarray}
where $\eta_{j}(\tilde{t})$ $(j=1, 2)$ denote independent Gaussian
white noises with zero means and correlation functions $\langle
\eta_{j}(\tilde{t}_{2}) \eta_{j}(\tilde{t}_{1}) \rangle = \delta
(\tilde{t}_{2} -\tilde{t}_{1})$, leads to the same Fokker-Planck
equation (\ref{F-Pfw}).\cite {DSTH2} This means that the above system
is stochastically equivalent to the stochastic Landau-Lifshitz
equation (\ref{L-L}). In what follows, we will use Eq.~(\ref{L-L2})
to numerically study the thermal stability of the precessional modes
of the magnetic moment $\mathbf{m}(t)$.

\subsection{Definition, boundary conditions, and transformation
properties of the lifetime} \label{Def}

At zero noise intensity, the rotating magnetic field $\mathbf{h}(t)$
can induce stable precessional modes of $\mathbf{m}(t)$ of two
types.\cite{BSM,BMS} In the first, P mode, the precession angle
$\Theta(t)$ is a constant and $\mathbf{m}(t)$ in the laboratory frame
is a periodic function of time. In the second, Q mode, the precession
angle varies periodically and $\mathbf{m}(t)$ becomes a
quasi-periodic function of time (because the periods of $\Theta(t)$
and $\mathbf{h}(t)$ are in general not commensurable). Some
properties of these modes related to the steady-state will be
considered in Sec.~\ref{Prec}. Here, we use only the fact that,
depending on the parameters $\tilde{h}, \tilde{\omega}, \rho,$ and
$\lambda$, one or two precessional modes may exist in steady state.
In the latter case, one mode occurs in the up state ($\sigma=+1$) and
the other in the down state ($\sigma=-1$) of $\mathbf{m}(t)$. We
assume that for a given set of the above parameters the magnetic
moment $\mathbf{m}(t)$ is in the state $\sigma$ if $m_{z}(t)$ tends
to $\sigma m$ as $\tilde{h}$ \textit{slowly} decreases to zero. It
should be noted that the last condition is important because a sharp
decrease of $\tilde{h}$ can switch $\mathbf{m}(t)$ to another state.

In steady-state, the reference modes are stable and transitions
between them are impossible. However, these transitions can occur
under thermal fluctuations. In this case the precessional modes
become metastable and the magnetic moment remains in a given state
$\sigma$ for some (dimensionless) random time $\tilde{t}_{\sigma}$.
The average value of this time, i.e., the lifetime $\mathcal
{T}_{\sigma}$ of the metastable state, can be determined using the
MFPT method.\cite{Ris,Gar} The basis of this method is the backward
Fokker-Planck equation (\ref{F-Pbw}), which should be written for a
given state $\sigma$. To this end, we add the index $\sigma$ to all
angle variables, replace the conditional probability density $P$ by
$P_{\sigma} = P_{\sigma} (\theta_{\sigma}, \psi_{\sigma}, \tilde{t}
|\theta'_{\sigma}, \psi'_{\sigma}, \tilde{t}')$, and assume that
$\theta_{+1}, \theta'_{+1} \in (0, \theta_{0})$, $\theta_{-1},
\theta'_{-1} \in (\pi - \theta_{0}, \pi)$ and $\psi_{\sigma},
\psi'_{\sigma} \in (0, 2\pi)$. Here, the angle $\theta_{0}$ ($\pi/2<
\theta_{0} <\pi$) should be chosen so that the time average of the
precession angle $\Theta_{ \sigma}(t)$ [we recall that $\Theta_{
\sigma} (t)$ for the P modes does not depend on $t$] is relatively
close to $\theta_{0}$ and to $\pi - \theta_{0}$ for $\sigma=-1$ and
+1, respectively. To meet these requirements, in our numerical
simulations we assume that $\theta_{0} = 0.8\pi$.

In accordance with the MFPT approach, we consider a circular cone
with the cone angle $\pi(1-\sigma)/2 + \sigma \theta_{0}$ as the
absorbing boundary for the magnetic moment in the $\sigma$ state,
i.e., $P_{\sigma} |_{\theta'_{\sigma} = \pi(1-\sigma)/2 + \sigma
\theta_{0}} = 0$. In this case, taking into account that $P_{\sigma}
= P_{\sigma} (\theta_{\sigma}, \psi_{\sigma}, u| \theta'_{ \sigma},
\psi'_{ \sigma}, 0)$ ($u = \tilde{t} - \tilde{t}'$), the lifetime can
be defined as $\mathcal {T}_{\sigma} = \int_{0}^{\infty}du
Q_{\sigma}$, where
\begin{equation}
    Q_{\sigma} =\int_{0}^{2\pi}d\psi_{\sigma} \int_{(\pi -
    \theta_{0}) (1-\sigma)/2}^{\pi(1 - \sigma)/2 +
    \theta_{0}(1 + \sigma)/2} d\theta_{\sigma} P_{\sigma}
    \label{defQ}
\end{equation}
is the probability that the magnetic moment stays in the state
$\sigma$ up to a given value of the time difference $u$. Finally,
using the relations $Q_{\sigma}|_{u=0} =1$ and $Q_{\sigma}
|_{u=\infty} =0$, one can make sure that the lifetime $\mathcal
{T}_{\sigma} = \mathcal{T}_{\sigma}(\theta'_{\sigma},
\psi'_{\sigma})$ is governed by the partial differential equation
\begin{eqnarray}
    &\displaystyle \frac{\partial^{2} \mathcal {T}_{\sigma}}
    {\partial\theta'^{2}_{\sigma}} + \frac{1}{\sin^{2}
    \theta'_{\sigma}}\frac{\partial^{2} \mathcal {T}_{\sigma}}
    {\partial\psi'^{2}_{\sigma}} + \bigg(\cot\theta'_{\sigma}
    + \frac{2 a}{\lambda} u(\theta'_{\sigma},\psi'_{\sigma})
    \bigg)\frac{\partial \mathcal {T}_{\sigma}}{\partial
    \theta'_{\sigma}}&
    \nonumber\\[6pt]
    & \displaystyle + \frac{2 a}{\lambda} [v(\theta'_{
    \sigma}, \psi'_{\sigma}) - \rho \tilde{\omega}]
    \frac{\partial \mathcal {T}_{\sigma}} {\partial\psi'_{
    \sigma}} = -\frac{2a}{\lambda}.&
    \label{MFPT1}
\end{eqnarray}

At the absorbing boundary, the solution of this equation must satisfy
the condition
\begin{equation}
    \mathcal {T}_{\sigma} \big|_{\theta_{\sigma}' =
    \pi (1-\sigma)/2 + \sigma\theta_{0}}  = 0.
    \label{absT}
\end{equation}
One more important property of the lifetime is that it is a finite
function of $\theta_{\sigma}'$ and $\psi_{\sigma}'$. In order to
prove this statement, let us first approximate the stochastic
dynamics of the magnetic moment by a random walk on the sphere
characterized by a dimensionless discrete time $\tilde{t} = n\tau$,
where $n = 0,1,\ldots$ and $\tau$ is the time step. Then, denoting
$r_{\sigma n}$ the probability that the magnetic moment stays in the
state $\sigma$ after the $n$th step, we can write $Q_{\sigma} =
\prod_{n=1}^{u/\tau} r_{\sigma n}$. If the maximal element of the set
$\{ r_{\sigma n} \}$ equals $R_{\sigma}$ then $Q_{\sigma
}<R_{\sigma}^{u/\tau}$ and, as a consequence, $\mathcal {T}_{
\sigma}< \int_{0} ^{\infty}du R_{\sigma}^{u/\tau} = \tau/|\ln
R_{\sigma}|$. Finally, taking into account that the condition
$\theta_{0}<\pi$ implies that $R_{\sigma}<1$, we obtain the desired
result: $\mathcal {T}_{\sigma}< \infty$. It should also be noted that
since the maximum angular distance to the absorbing boundary occurs
at $\theta_{\sigma}' = \pi(1 - \sigma)/2$, i.e., $\max{\mathcal
{T}_{\sigma}} = \mathcal {T}_{\sigma} \big|_{\theta_{\sigma}' = \pi(1
- \sigma)/2}$, the condition of finiteness of the lifetime can be
written in the form
\begin{equation}
    \mathcal {T}_{\sigma} \big|_{\theta_{\sigma}' = \pi(1 -
    \sigma)/2}  < \infty.
    \label{finT}
\end{equation}

The above result shows the importance of knowing the solution of
Eq.~(\ref{MFPT1}) in a small vicinity of the point $\theta_{\sigma}'
= \pi(1 - \sigma)/2$. Assuming that $\theta_{\sigma}' = \pi(1 -
\sigma)/2 + \sigma \xi_{\sigma}$ ($\xi_{ \sigma}>0$), this equation
at $\xi_{ \sigma} \to 0$ reduces to
\begin{eqnarray}
    &\displaystyle \frac{\partial^{2} \mathcal {T}_{\sigma}}
    {\partial\xi^{2}_{\sigma}} + \frac{1}{\xi_{\sigma}^{2}}
    \frac{\partial^{2} \mathcal {T}_{\sigma}}
    {\partial\psi'^{2}_{\sigma}} + \frac{1}{\xi_{\sigma}}
    \frac{\partial \mathcal {T}_{\sigma}}{\partial\xi_{\sigma}}
    -\frac{2a\tilde{h}}{\lambda \xi_{\sigma}}
    (\sigma \cos\psi'_{\sigma}  &
    \nonumber\\[6pt]
    & \displaystyle + \lambda \sin \psi'_{\sigma})
    \frac{\partial \mathcal {T}_{\sigma}} {\partial
    \psi'_{\sigma}} = -\frac{2a}{\lambda}.&
    \label{AsEq_MFPT1}
\end{eqnarray}
Its solution can be represented as
\begin{equation}
    \mathcal{T}_{\sigma} = c \ln\xi_{\sigma} + f_{0} +
    \sum_{l=1}^{\infty} f_{l}(\psi'_{\sigma})\xi_{\sigma}^{l},
    \label{Sol_AsEq}
\end{equation}
where $c$ and $f_{0}$ are constant parameters, and the functions
$f_{l} =f_{l} (\psi'_{\sigma})$ satisfy the ordinary differential
equations $d^{2}f_{1} /d\psi'^{2}_{\sigma} + f_{1} =0$ and
\begin{equation}
    \frac{d^{2}f_{l}}{d\psi'^{2}_{\sigma}} + l^{2}f_{l} -
    \frac{2a\tilde{h}}{\lambda}(\sigma \cos\psi'_{\sigma}
    + \lambda \sin \psi'_{\sigma})\frac{df_{l-1}}{d\psi'_{\sigma}}
    = - \frac{2a}{\lambda} \delta_{l2}
    \label{f_{l}}
\end{equation}
with $l\geq 2$ and $\delta_{ln}$ being the Kronecker delta. According
to Eq.~(\ref{Sol_AsEq}), the finiteness condition (\ref{finT}) takes
the form $c=0$, which in turn is equivalent to $\lim_{\xi_{\sigma}
\to 0} \xi_{\sigma} \partial \mathcal{T}_{ \sigma}/\partial
\xi_{\sigma} =0$. We note that at $c=0$ the derivative $\partial
\mathcal{T}_{\sigma}/\partial \theta_{\sigma}'$ in the point
$\theta_{\sigma}' = \pi(1 - \sigma)/2$, in general, does not vanish:
$\partial \mathcal{T}_{\sigma}/\partial \theta_{\sigma}'|
_{\theta_{\sigma}' =\pi(1 - \sigma)/2} = \sigma f_{1}$. However, if
$\tilde{h} =0$ then $f_{1}=0$ (this is so because in this case
$\mathcal {T}_{\sigma}$ does not depend on $\psi'_{ \sigma}$), and
the finiteness condition (\ref{finT}) reduces to the reflecting
boundary condition
\begin{equation}
    \frac{\partial\mathcal {T}_{\sigma}} {\partial\theta'_{\sigma}}
    \bigg|_{\theta_{\sigma}' =\pi(1 - \sigma)/2} =0.
    \label{reflT}
\end{equation}
It should also be noted that at $\tilde{h} \ne 0$ the same reflecting
boundary condition holds for the average lifetime $\overline{\mathcal
{T}}_{\sigma} = (1/2\pi) \int_{0}^{2\pi} d\psi'_{\sigma} \mathcal
{T}_{\sigma}$, since $\overline{f}_{1} =0$.

Now, using the general equation (\ref{MFPT1}), we can establish the
transformation properties of its solution $\mathcal {T}_{\sigma} =
\mathcal {T}_{\sigma} (\theta'_{\sigma}, \psi'_{\sigma}; \rho)$ (for
clarity, the dependence of $\mathcal {T}_{\sigma}$ on $\rho$ is shown
explicitly), which is assumed to obey the conditions (\ref{absT}) and
(\ref{finT}). Toward this end, let us introduce the change of
variables
\begin{equation}
    \theta'_{\sigma} = \pi - \theta'_{-\sigma},  \qquad
    \psi'_{\sigma} = 2\pi - \psi'_{-\sigma}.
    \label{change}
\end{equation}
Taking into account that $u(\theta'_{\sigma}, \psi'_{\sigma}) =
-u(\theta' _{-\sigma}, \psi'_{-\sigma})$ and $u(\theta'_{\sigma},
\psi'_{\sigma}) = -u(\theta'_{-\sigma}, \psi'_{-\sigma})$, one can
make sure that Eq.~(\ref{MFPT1}) in the new variables $\theta'_{-
\sigma}$ and $\psi'_{-\sigma}$ becomes
\begin{equation}
    \begin{array}{c}
    \displaystyle \frac{\partial^{2} \mathcal {H}
    _{\sigma}} {\partial\theta'^{2}_{-\sigma}} +
    \frac{1}{\sin^{2} \theta'_{-\sigma}}\frac{\partial^{2}
    \mathcal {H}  _{\sigma}} {\partial\psi'^{2}_{-\sigma}} +
    \bigg(\! \cot  \theta'_{-\sigma} + \frac{2 a}{\lambda}
    u(\theta'_{-\sigma}, \psi'_{-\sigma})\bigg) \\[16pt]
    \displaystyle \times \frac{\partial \mathcal {H}
    _{\sigma}} {\partial\theta'_{-\sigma}} + \frac{2 a}
    {\lambda} [v(\theta'_{-\sigma}, \psi'_{-\sigma}) +
    \rho \tilde{\omega}] \frac{\partial \mathcal {H}_{\sigma}}
    {\partial \psi'_{-\sigma}} = -\frac{2a}{\lambda},
    \end{array}
    \label{MFPT2}
\end{equation}
where $\mathcal {H}_{\sigma} = \mathcal {T}_{\sigma} (\pi -
\theta'_{-\sigma}, 2\pi - \psi'_{-\sigma}; \rho)$. In accordance with
the transforms (\ref{change}), the conditions (\ref{absT}) and
(\ref{finT}) for $ \mathcal {H}_{ \sigma}$ take the form $\mathcal
{H}_{\sigma}|_{ \theta_{-\sigma}' = \pi (1 +\sigma)/2 - \sigma
\theta_{0}}  = 0$ and $\mathcal {H}_{ \sigma}|_{ \theta_{-\sigma}' =
\pi(1 + \sigma)/2} < \infty$, respectively. Therefore, comparing
Eqs.~(\ref{MFPT1}) and (\ref{MFPT2}) and the corresponding absorbing
and finiteness conditions, one can conclude that $\mathcal
{H}_{\sigma}$ is equal to $\mathcal {T}_{-\sigma} (\theta'_{-
\sigma}, \psi'_{-\sigma}; -\rho)$, i.e.,
\begin{equation}
    \mathcal {T}_{\sigma}(\theta'_{\sigma}, \psi'_{\sigma};\rho)
    = \mathcal {T}_{-\sigma}(\pi - \theta'_{\sigma},
    2\pi - \psi'_{\sigma};-\rho).
    \label{transf1}
\end{equation}
For the average lifetime $\overline\mathcal {T}_{\sigma}$ this
transformation property reads: $\overline{\mathcal {T}}_{\sigma}(
\theta'_{\sigma};\rho) = \overline{\mathcal {T}}_{-\sigma}(\pi -
\theta'_{\sigma};-\rho)$.

In general, the arguments $\theta'_{\sigma}$ and $\psi'_{\sigma}$ of
$\mathcal {T}_{\sigma}$ are arbitrary and can be properly chosen to
best suit the problem. In this paper we are interested in the
lifetime of the precessional modes \textit{reaching} the steady
state. Therefore, the angles $\theta'_{ \sigma}$ and $\psi'_{\sigma}$
should be associated with the solution of deterministic (when
$a=\infty$) Landau-Lifshitz equations (\ref{L-L2}) at some time
$\tilde{t} = \tilde{t}_{\mathrm{st}}$, i.e., $\theta'_{ \sigma} =
\theta_{ \sigma} (\tilde{t}_{\mathrm{st}})$ and $\psi'_{\sigma} =
\psi_{\sigma} (\tilde{t}_{ \mathrm{st}})$. To be sure that the
magnetic moment is near the steady state, we assume that
$\tilde{t}_{\mathrm{st}} > \tilde{t}_{0} + \tilde{t}_{ \mathrm{
rel}}$, where $\tilde{t}_{0}$ is the time of increasing  the rotating
field amplitude from $0$ to a given value $\tilde{h}$, and $\tilde{
t}_{ \mathrm{rel}} = 2/\lambda$ is the relaxation time. It should be
emphasized that the time $t_{0}$ must be chosen large enough to
prevent the dynamical switching from the state $\sigma$ to the state
$-\sigma$. In the P mode, the angles $\theta_{ \sigma} (\tilde{t}
_{\mathrm{st}})$ and $\psi_{\sigma} (\tilde{t}_{ \mathrm{st}})$
approach the limiting precession angle $\Theta_{ \sigma} = \lim_{
\tilde{t} \to \infty} \Theta_{\sigma} (\tilde{t})$ and the limiting
difference of phases $\Psi_{\sigma} = \lim_{ \tilde{t} \to \infty}
\Psi_{\sigma} (\tilde{t})$, respectively. As a consequence, the
lifetime of this mode
\begin{equation}
    \mathcal {T}_{\sigma} = \mathcal {T}_{\sigma}
    (\Theta_{\sigma}, \Psi_{\sigma};\rho)
    \label{Pmod}
\end{equation}
does not depend on the precise choice of $\tilde{t}_{ \mathrm{st}}$.
In contrast, in the Q mode the precession angle $\Theta_{\sigma}
(\tilde{t})$ is a periodic function of time $\tilde{t}$ with a period
$\widetilde{T}_{ \mathrm{Q}}$ and the difference of phases is given
by $\Psi_{\sigma} (\tilde{t}) = -\nu \tilde{t} + \Phi_{\sigma}
(\tilde{t})$, where $\nu\geq 0$ and $\Phi_{\sigma} (\tilde{t})$ is
also a periodic function with the same period $\widetilde{T}_{
\mathrm{Q}}$ (see Sec.~\ref{Prec}). Thus the lifetime of the Q mode,
in general, depends on $\tilde{t}_{\mathrm{st}}$:
\begin{equation}
    \mathcal {T}_{\sigma} = \mathcal {T}_{\sigma}(\Theta_{
    \sigma} (\tilde{t}_{\mathrm{st}}),\Psi_{\sigma}
    (\tilde{t}_{\mathrm{st}});\rho).
    \label{Qmod}
\end{equation}
However, at $a\gg 1$ this dependence is very weak and can be safely
neglected (see Sec.~\ref{Simulat}).

In the case of P mode the limiting angles $\Theta_{\sigma}$ and
$\Psi_{\sigma}$ depend also on the parameters $\tilde{h}$,
$\tilde{\omega}$, $\lambda$ and $\rho$. However, for brevity, we keep
only the parameter $\rho$, i.e., $\Theta_{\sigma} = \Theta_{\sigma}
[\rho]$ and $\Psi_{\sigma} = \Psi_{\sigma} [\rho]$, which together
with the state parameter $\sigma$ describes the transformation
properties of these angles. To find them, we use Eq.~(\ref{def uv})
for representing the stationary Landau-Lifshitz equations
$u(\Theta_{\sigma} [\rho], \Psi_{\sigma}[\rho])=0$ and $v(\Theta_{
\sigma} [\rho], \Psi_{\sigma} [\rho]) = \rho \tilde{ \omega}$ as
\begin{eqnarray}
    &\displaystyle \cos \Psi_{\sigma}(\rho) =
    \frac{1}{\tilde{h}}(\sin \Theta_{\sigma}
    [\rho] - \rho \kappa \tan\Theta_{\sigma}[\rho]),&
    \nonumber\\[3pt]
    & \displaystyle \sin \Psi_{\sigma}[\rho] =-
    \frac{\rho \lambda \kappa}{\tilde{h}}
    \sin \Theta_{\sigma}[\rho],&
    \label{st1}
\end{eqnarray}
where $\kappa = \tilde{\omega}/(1+\lambda^{2})$. From these
equations, it is straightforward to obtain the desired result:
\begin{equation}
    \Theta_{\sigma}[\rho] = \pi - \Theta_{-\sigma}[-\rho],
    \qquad
    \Psi_{\sigma}[\rho] = 2\pi - \Psi_{-\sigma}[-\rho].
    \label{sym_sol}
\end{equation}
Because the transformation properties (\ref{sym_sol}) are similar to
those in Eq.~(\ref{change}), from Eq.~(\ref{transf1}) one gets the
transformation property of the lifetime of the P mode
\begin{equation}
    \mathcal {T}_{\sigma}(\Theta_{\sigma}[\rho],\Psi_{
    \sigma}[\rho];\rho) = \mathcal {T}_{-\sigma}
    (\pi - \Theta_{\sigma}[\rho], 2\pi -
    \Psi_{\sigma}[\rho];-\rho).
    \label{transf2}
\end{equation}
It shows that the lifetimes characterized by the pairs $\{ \sigma,
\rho \}$ and $\{ -\sigma, -\rho \}$ are the same.

\section{ANALYTICAL SOLUTION OF Eq.~(\ref{MFPT1})}
\label{AnRes}
\subsection{Three-mode approximation}
\label{One-mode}

The analytical determination of the lifetimes of the precessional
modes implies the solution of Eq.~(\ref{MFPT1}) with the absorbing
and finiteness conditions (\ref{absT}) and (\ref{finT}). Since the
lifetime $\mathcal{T}_{ \sigma}$ is a periodic function of
$\psi'_{\sigma}$ with the period $2\pi$, it can be expressed as the
Fourier series
\begin{equation}
    \mathcal{T}_{\sigma} =  \sum_{n=-\infty}^{\infty}
    \mathcal{T}_{\sigma n} (\theta'_{\sigma})
    e^{in\psi'_{\sigma}}.
    \label{SolT}
\end{equation}
To guarantee the reality of $\mathcal{T}_{ \sigma}$, we assume that
the coefficients $\mathcal{T}_{\sigma n} = \mathcal{T}_{ \sigma n}
(\theta'_{ \sigma})$ of the series satisfy the condition
$\mathcal{T}_{\sigma -n} = \mathcal{T}_{ \sigma n}^{*}$ (the asterisk
denotes complex conjugation). Substituting this series into
Eq.~(\ref{MFPT1}) and introducing the differential operators
\begin{eqnarray}
    \hat{L}_{n} \!&=&\! \frac{d^{2}}{d\theta'^{2}_{\sigma}} +
    (\cot \theta'_{\sigma} -a \sin 2\theta'_{\sigma})
    \frac{d}{d\theta'_{\sigma}} - \frac{n^{2}}{\sin^{2}
    \theta'_{\sigma}}
    \nonumber\\[4pt]
    &&\! + \, i\frac{2an}{\lambda} (\cos \theta'_{\sigma} -
    \rho \tilde{\omega}),
    \nonumber\\[4pt]
    \hat{N}_{n} \!&=&\! (\lambda \cos \theta'_{\sigma} - i)
    \frac{d}{d\theta'_{\sigma}} + \frac{\lambda n}
    {\sin \theta'_{\sigma}} - i n \cot \theta'_{\sigma},
    \label{L,N}
\end{eqnarray}
one obtains for $\mathcal{T}_{\sigma n}$ an infinite set of coupled
equations
\begin{equation}
    \hat{L}_{n}\mathcal{T}_{\sigma n} + \frac{a \tilde{h}}{\lambda}
    \big( \hat{N}_{n+1}\mathcal{T}_{\sigma n+1} +
    \hat{N}_{-n+1}^{*}\mathcal{T}_{\sigma n-1} \big) =
    -\frac{2a}{\lambda} \delta_{n0}.
    \label{Tn}
\end{equation}
Like $\mathcal{T}_{\sigma}$, the coefficients $\mathcal{T}_{\sigma
n}$ must satisfy both the absorbing boundary condition $\mathcal
{T}_{\sigma n} |_{ \theta_{\sigma}' = \pi (1-\sigma)/2 +
\sigma\theta_{0}}  = 0$ and the finiteness condition $\mathcal
{T}_{\sigma n} |_{\theta_{\sigma}' = \pi(1 - \sigma)/2}  < \infty$.

In the three-mode approximation, when $\mathcal{T}_{\sigma n} = 0$
for all $|n| \geq 2$, an infinite set of equations (\ref{Tn}) reduces
to a set of three equations for $\mathcal{T}_{\sigma 0}$,
$\mathcal{T}_{\sigma 1}$ and $\mathcal{T}_{\sigma 1}^{*}$. Since
$\mathcal{T}_{\sigma}$ is real, it is convenient to consider instead
of $\mathcal{T}_{ \sigma 1}$ and $\mathcal{T}_{ \sigma 1}^{*}$ the
real and imaginary parts of $\mathcal{T}_{ \sigma 1}$, i.e.,
$\mathcal{T}_{ \sigma 1}^{+} = \mathrm{Re}\, \mathcal{T}_{ \sigma 1}$
and $\mathcal{T}_{ \sigma 1}^{-} = \mathrm{Im}\,\mathcal{T}_{ \sigma
1}$. Then, in this approximation, the lifetime (\ref{SolT}) is given
by
\begin{equation}
    \mathcal{T}_{\sigma} = \mathcal{T}_{\sigma 0} +
    2 \mathcal{T}_{\sigma 1}^{+} \cos \psi'_{\sigma} -
    2 \mathcal{T}_{\sigma 1}^{-} \sin \psi'_{\sigma}
    \label{appr}
\end{equation}
and, denoting the real and imaginary parts of an operator $\hat{O}$
as $\hat{O}^{+}$ and $\hat{O}^{-}$, respectively, from
Eqs.~(\ref{Tn}) and (\ref{L,N}) one obtains a set of three coupled
equations for $\mathcal{T}_{\sigma 0}$, $\mathcal{T}_{\sigma 1}^{+}$
and $\mathcal{T}_{\sigma 1}^{-}$:
\begin{eqnarray}
    &\displaystyle\hat{L}_{0}\mathcal{T}_{\sigma 0} +
    \frac{2a}{\lambda} \big(1 + \tilde {h}\hat{N}_{1}^{+}
    \mathcal{T}_{\sigma 1}^{+} - \tilde {h}\hat{N}_{1}^{-}
    \mathcal{T}_{\sigma 1}^{-}  \big) = 0,&
    \label{eqT0} \\[4pt]
    &\displaystyle \hat{L}_{1}^{+} \mathcal{T}_{\sigma 1}^{+}
    - \hat{L}_{1}^{-} \mathcal{T}_{\sigma 1}^{-} =
    - \frac{a \tilde{h}}{\lambda} \hat{N}_{0}^{+}
    \mathcal{T}_{\sigma 0},&
    \label{eqT+} \\[4pt]
    &\displaystyle \hat{L}_{1}^{+} \mathcal{T}_{\sigma 1}^{-}
    + \hat{L}_{1}^{-} \mathcal{T}_{\sigma 1}^{+} =
    \frac{a \tilde{h}}{\lambda} \hat{N}_{0}^{-}
    \mathcal{T}_{\sigma 0}.&
    \label{eqT-}
\end{eqnarray}

Using Eq.~(\ref{eqT0}), the function $\mathcal{T}_{\sigma 0}
(\theta'_{ \sigma})$ can be expressed through $\mathcal{T}_{\sigma
1}^{+} (\theta'_{ \sigma})$ and $\mathcal{T}_{\sigma 1}^{-}
(\theta'_{ \sigma})$. Indeed, considering
\begin{equation}
    F_{\sigma}(\theta'_{\sigma}) = \tilde {h}\hat{N}_{1}^{+}
    \mathcal{T}_{\sigma 1}^{+} - \tilde {h}\hat{N}_{1}^{-}
    \mathcal{T}_{\sigma 1}^{-}
    \label{defF}
\end{equation}
as a given function of $\theta'_{ \sigma}$, one can write a formal
solution of Eq.~(\ref{eqT0}).\cite{PZ} From this solution, by
satisfying the absorbing and reflecting boundary conditions $\mathcal
{T}_{\sigma 0} |_{ \theta_{\sigma}' = \pi (1-\sigma)/2 +
\sigma\theta_{0}}  = 0$ and $d\mathcal {T}_{\sigma 0}/
d\theta'_{\sigma} |_{\theta_{\sigma}' =\pi(1 - \sigma)/2} =0$ (since
$\mathcal {T}_{\sigma 0}$ does not depend on $\theta_{\sigma}'$, the
last condition is equivalent to the finiteness condition $\mathcal
{T}_{\sigma 0} |_{\theta_{ \sigma}' = \pi(1 - \sigma)/2}  < \infty$,
see Sec.~\ref{Def}), we obtain
\begin{equation}
    \mathcal {T}_{\sigma 0} = \frac{2a}{\lambda} \int_{\cos
    \theta_{0}}^{\sigma \cos \theta_{\sigma}'}\! dx
    \frac{e^{-ax^{2}}}{1 - x^{2}} \int_{x}^{1}\!dy [1 + F_{\sigma}
    (\arccos \sigma y)] e^{ay^{2}}.
    \label{T0}
\end{equation}
Substituting this result into Eqs.~(\ref{eqT+}) and (\ref{eqT-}), one
can readily get the coupled integro-differential equations for
$\mathcal{T}_{\sigma 1}^{+}$ and $\mathcal{T}_{\sigma 1}^{-}$ which,
however, are too complicated to be solved in the general case.
Moreover, these equations are not closed because, in accordance with
Eqs.~(\ref{defF}), (\ref{eqT+}) and (\ref{eqT-}), the function
$F_{\sigma} (\arccos \sigma y)$ depends on $\mathcal {T}_{\sigma 0}$.
Fortunately, an important case of high anisotropy barrier can be
studied in detail.

\subsection{High anisotropy barrier}
\label{High}

In the case of high anisotropy barrier (or low temperatures), when
the condition $a=mH_{a}/2k_{B}T \gg 1$ holds, the main contribution
to the first integral in Eq.~(\ref{T0}) comes either from a small
vicinity of the point $x=0$ (if $\theta_{\sigma}'$ is not too close
to $\pi/2$ and $\sigma \cos \theta_{ \sigma}'>0$) or from the
vicinity of the point $x= \sigma \cos \theta_{ \sigma}'$ (if $\sigma
\cos \theta_{ \sigma}'<0$). It should be noted that the latter
situation can be realized only in the case of Q mode with $\max
\Theta_{+1} (\tilde{t})> \pi/2$ or $\min \Theta_{-1} (\tilde{t})<
\pi/2$. Here, we restrict our theoretical analysis by considering
small amplitudes of the rotating field that do not exceed the
threshold amplitude of the Q mode. Such rotating field can induce
only the P mode and, as a consequence, the former situation is always
realized (see also Sec.~\ref{Prec}). Therefore, taking into account
that in this case the integrals $\int_{\cos \theta_{0}}^{ \sigma \cos
\theta_{\sigma}'} dx e^{-ax^{2}}$ and $\int_{0}^{1} dy e^{ay^{2}}$ at
$a \gg 1$ can be approximated by $\int_{ -\infty}^{\infty} dx
e^{-ax^{2}} = \sqrt{\pi/a}$ and $e^{a}/2a$, respectively,
Eq.~(\ref{T0}) reduces to
\begin{equation}
    \mathcal {T}_{\sigma 0} = \frac{e^{a}}{\lambda}
    \sqrt{\frac{\pi}{a}}\left( 1 + 2ae^{-a}\int_{0}^{1}dy
    F_{\sigma} (\arccos \sigma y)e^{ay^{2}} \right).
    \label{T02}
\end{equation}

In order to evaluate the integral in Eq.~(\ref{T02}), we note that
small vicinities of the limits of the integration determine the
asymptotic behavior of this integral at $a \to \infty$. More
precisely, the lower limit is responsible for the high-frequency
behavior of this integral and the upper one for its behavior in the
resonant case. Hereafter, we call the rotating field resonant if
$\tilde{\omega} \sim 1$ and the direction of its rotation coincides
with the direction of the natural precession of the magnetic moment,
i.e., if $\sigma \rho = +1$. To study these cases in a unified way,
it is convenient to write the integral in Eq.~(\ref{T02}) as a sum of
two terms, $F_{\sigma} (\arccos \sigma) e^{a}/2a$ and $F_{\sigma}
(\arccos 0)$, which correspond to the resonant and high frequencies,
respectively. Thus, taking into account that $\arccos \sigma = \pi(1-
\sigma)/2$ and $\arccos 0 = \pi/2$, Eq.~(\ref{T02}) yields
\begin{equation}
    \mathcal {T}_{\sigma 0} = \frac{e^{a}}{\lambda}
    \sqrt{\frac{\pi}{a}}\left[ 1 + F_{\sigma}\!
    \left( \frac{\pi}{2}(1-\sigma) \right) + 2ae^{-a}
    F_{\sigma}\! \left( \frac{\pi}{2} \right) \right].
    \label{AsT0}
\end{equation}
It is important to stress at this point that Eq.~(\ref{AsT0}) is not
an exact asymptotic formula for $\mathcal {T}_{\sigma 0}$ because the
terms $F_{\sigma} (\pi(1- \sigma)/2)$ and $2a e^{-a}F_{\sigma}
(\pi/2)$ correspond to different frequencies.

\subsubsection{Vicinity of the point $\theta_{\sigma}'=\pi(1-
\sigma)/2$} \label{Vic1}

To calculate  $F_{\sigma} (\pi(1- \sigma)/2)$, we need to solve
Eqs.~(\ref{eqT+}) and (\ref{eqT-}) in the vicinity of the point
$\theta_{ \sigma}' =\pi(1- \sigma)/2$. Assuming that $\theta_{
\sigma}' = \pi(1- \sigma)/2 + \sigma \eta_{\sigma}$ ($0<\eta_{\sigma}
\ll 1$) and $a\tilde{h} \ll 1$, these equations can be rewritten as
\begin{eqnarray}
    &\displaystyle \hat{L}_{1}^{+}\mathcal{T}_{\sigma 1}^{+}
    - \frac{2a}{\lambda} (\sigma - \rho \tilde{\omega})
    \mathcal{T}_{\sigma 1}^{-} = -a\tilde{h}
    \frac{d\mathcal{T}_{\sigma 0}} {d\eta_{\sigma}},&
    \nonumber \\[4pt]
    &\displaystyle \hat{L}_{1}^{+}\mathcal{T}_{\sigma 1}^{-}
    + \frac{2a}{\lambda} (\sigma - \rho \tilde{\omega})
    \mathcal{T}_{\sigma 1}^{+} = -\sigma
    \frac{a\tilde{h}}{\lambda}\frac{d\mathcal{T}_{\sigma 0}}
    {d\eta_{\sigma}},&
    \label{eqsT+-a}
\end{eqnarray}
where
\begin{equation}
    \hat{L}_{1}^{+} = \frac{d^{2}} {d\eta_{\sigma}^{2}} +
    \bigg( \frac{1}{\eta_{\sigma}} - 2a\eta_{\sigma} \bigg)
    \frac{d} {d\eta_{\sigma}} - \frac{1}{\eta^{2}_{\sigma}}.
    \label{L0a}
\end{equation}
Since the approximate formula (\ref{AsT0}) does not depend on
$\theta_{ \sigma}'$, it cannot be used to determine the derivative
$d\mathcal {T}_{ \sigma 0}/d\eta_{\sigma}$. Therefore we use an exact
result:
\begin{equation}
    \frac{d\mathcal {T}_{\sigma 0}}{d\theta_{\sigma}'} =
    - \sigma \frac{2ae^{-a\cos^{2}\!\theta_{\sigma}'}}
    {\lambda \sin\theta_{\sigma}'} \int_{\sigma \cos
    \theta_{\sigma}'}^{1}\! dy [1 + F_{\sigma}
    (\arccos \sigma y)] e^{ay^{2}},
    \label{derT0}
\end{equation}
following from Eq.~(\ref{T0}), which at $\eta_{\sigma} \ll 1$ gives
\begin{equation}
    \frac{d\mathcal {T}_{\sigma 0}}{d\eta_{\sigma}} =
    - \frac{a}{\lambda}\! \left[ 1 + F_{\sigma}\!
    \left( \frac{\pi}{2}(1-\sigma) \right)\right]\!
    \eta_{\sigma}.
    \label{T0a}
\end{equation}

It is not difficult to verify that an exact solution of
Eq.~(\ref{eqsT+-a}) that vanishes as $\tilde{h} \to 0$ has the form
\begin{eqnarray}
    &\displaystyle \mathcal{T}_{\sigma 1}^{+} = -\frac{(1 -
    \lambda^{2} -\sigma\rho \tilde{\omega})\tilde{h}}{2[(1 -
    \sigma\rho \tilde{\omega})^{2} +\lambda^{2}]}
    \frac{d\mathcal{T}_{\sigma 0}} {d\eta_{\sigma}},&
    \nonumber \\[4pt]
    &\displaystyle \mathcal{T}_{\sigma 1}^{-} = \sigma
    \frac{\lambda (2-\sigma\rho\tilde{\omega}) \tilde{h}}
    {2[(1-\sigma\rho \tilde{\omega})^{2} + \lambda^{2}]}
    \frac{d\mathcal{T}_{\sigma 0}} {d\eta_{\sigma}},&
    \label{solT+-}
\end{eqnarray}
where $d\mathcal {T}_{ \sigma 0}/d\eta_{\sigma}$ is given by
Eq.~(\ref{T0a}). According to this equation, the solution
(\ref{solT+-}) contains an unknown parameter $F_{\sigma}(
\pi(1-\sigma)/2)$, which can be determined from the fitting condition
$F_{\sigma}( \pi(1-\sigma)/2) = ( \tilde {h}\hat{N}_{1}^{+}
\mathcal{T}_{\sigma 1}^{+} - \tilde {h}\hat{N}_{1}^{-}
\mathcal{T}_{\sigma 1}^{-}) |_{\eta_{\sigma} = 0}$ [see
Eq.~(\ref{defF})]. Taking into account that $\hat{N}_{1}^{+} =
\lambda d/d\eta_{\sigma} + \lambda/\eta_{\sigma}$ and
$\hat{N}_{1}^{-} = -\sigma d/d\eta_{\sigma} - \sigma/\eta_{\sigma}$,
this condition reduces to
\begin{equation}
    F_{\sigma}\! \left( \frac{\pi}{2}(1-\sigma) \right) =
    \frac{\lambda (1 + \lambda^{2}) \tilde{h}^{2}}
    {(1-\sigma\rho \tilde{\omega})^{2} + \lambda^{2}}
    \frac{d^{2}\mathcal{T}_{\sigma 0}} {d\eta_{\sigma}^{2}}.
    \label{fitt}
\end{equation}
Finally, substituting Eq.~(\ref{T0a}) into Eq.~(\ref{fitt}) and
imposing the commonly used condition $\lambda^{2} \ll 1$, one gets
\begin{equation}
    F_{\sigma}\! \left( \frac{\pi}{2}(1-\sigma) \right) =
    - \frac{a \tilde{h}^{2}}{(1-\sigma\rho \tilde{\omega})^{2}
    + \lambda^{2} + a\tilde{h}^{2}}.
    \label{F1}
\end{equation}

\subsubsection{Vicinity of the point $\theta_{\sigma}'=\pi/2$}
\label{Vic2}

To find $F_{\sigma}(\pi/2)$, we assume that $\theta_{\sigma}' = \pi/2
+ \xi_{\sigma}$ ($|\xi_{\sigma}| \ll 1$) and, as before, $a\tilde{h}
\ll 1$. Since in this case $\hat{L}_{1}^{-} = -2a(\xi_{ \sigma} +
\rho \tilde{\omega})/ \lambda$, $\hat{N}_{0}^{+} = -\lambda
\xi_{\sigma} d/d\xi_{\sigma}$ and $\hat{N}_{0}^{-} = -d/d\xi_{
\sigma}$, Eqs.~(\ref{eqT+}) and (\ref{eqT-}) take the form
\begin{eqnarray}
    &\displaystyle \hat{L}_{1}^{+}\mathcal{T}_{\sigma 1}^{+}
    +\frac{2a}{\lambda}(\xi_{\sigma} + \rho\tilde{\omega})
    \mathcal{T}_{\sigma 1}^{-} =  a\tilde{h} \xi_{\sigma}
    \frac{d\mathcal{T}_{\sigma 0}}{d\xi_{\sigma}},&
    \nonumber \\[4pt]
    &\displaystyle \hat{L}_{1}^{+}\mathcal{T}_{\sigma 1}^{-}
    - \frac{2a}{\lambda}(\xi_{\sigma} + \rho\tilde{\omega})
    \mathcal{T}_{\sigma 1}^{+} = -\frac{a\tilde{h}}
    {\lambda}\frac{d\mathcal{T}_{\sigma 0}}{d\xi_{\sigma}},&
    \label{eqsT+-b}
\end{eqnarray}
where
\begin{equation}
    \hat{L}_{1}^{+} = \frac{d^{2}} {d\xi_{\sigma}^{2}} +
    2a\xi_{\sigma}\frac{d} {d\xi_{\sigma}} - 1
    \label{L0b}
\end{equation}
and, as it follows from Eq.~(\ref{derT0}), the derivative $d\mathcal
{T}_{\sigma 0}/d\xi_{\sigma}$ at $\xi_{\sigma} \to 0$ is given by
\begin{equation}
    \frac{d\mathcal {T}_{\sigma 0}}{d\xi_{\sigma}} =
    - \sigma \frac{e^{a}}{\lambda}\left[ 1 + F_{\sigma}\!
    \left( \frac{\pi}{2}(1-\sigma) \right) + 2ae^{-a}
    F_{\sigma}\! \left( \frac{\pi}{2} \right) \right].
    \label{T0b}
\end{equation}

In general, the solution of Eq.~(\ref{eqsT+-b}) can be represented in
the form of the Taylor series: $\mathcal{T}_{\sigma 1}^{\pm} =
\sum_{n=0}^{\infty} c_{\sigma n}^{\pm} \xi_{\sigma}^{n}$. However,
since the main quantity of our interest is $F_{\sigma} (\pi/2) =
(\tilde {h}\hat{N}_{1}^{+} \mathcal{T}_{ \sigma 1}^{+} - \tilde
{h}\hat{N}_{1}^{-} \mathcal{T}_{\sigma 1}^{-})|_{ \xi_{\sigma}=0} =
\lambda \tilde{h} c_{\sigma 0}^{+} + \tilde{h} c_{\sigma 1}^{-}$
(here, $\hat{N}_{1}^{+} = \lambda$ and $\hat{N }_{1}^{-} = - d/d\xi_{
\sigma}$), we can restrict ourselves to the linear approximation,
i.e., $\mathcal{T}_{\sigma 1}^{\pm} = c_{\sigma 0}^{\pm} + c_{\sigma
1}^{\pm} \xi_{\sigma}$. Assuming that $\tilde{\omega} \gg \lambda/
2a$, after straightforward calculations one obtains
\begin{equation}
    c_{\sigma 0}^{-} = \frac{\lambda \tilde{h}}{4a\tilde{
    \omega}^{2}} \frac{d\mathcal{T}_{\sigma 0}}{d\xi_{\sigma}},
    \quad c_{\sigma 0}^{+} = \rho\frac{\tilde{h}}{2\tilde{
    \omega}} \frac{d\mathcal{T}_{\sigma 0}}{d\xi_{\sigma}}
    \label{c0}
\end{equation}
and
\begin{equation}
    c_{\sigma 1}^{-} = \rho\frac{\lambda(1+ \tilde{\omega}^{2})
    \tilde{h}} {2\tilde{\omega}(\lambda^{2} + \tilde{\omega}^{2})}
    \frac{d\mathcal{T}_{\sigma 0}} {d\xi_{\sigma}}, \quad
    c_{\sigma 1}^{+} = -\frac{\tilde{h}} {2(\lambda^{2} +
    \tilde{\omega}^{2})} \frac{d\mathcal{T}_{\sigma 0}}
    {d\xi_{\sigma}}.
    \label{c1}
\end{equation}
Therefore, the fitting condition $F_{\sigma} (\pi/2) = \lambda
\tilde{h} c_{\sigma 0}^{+} + \tilde{h} c_{\sigma 1}^{-}$ can be
reduced to the following one:
\begin{equation}
    F_{\sigma}\! \left( \frac{\pi}{2} \right) = \rho
    \frac{\lambda(1+ \lambda^{2} + 2\tilde{\omega}^{2})
    \tilde{h}^{2}} {2\tilde{\omega}(\lambda^{2} +
    \tilde{\omega}^{2})} \frac{d\mathcal{T}_{\sigma 0}}
    {d\xi_{\sigma}},
    \label{F2a}
\end{equation}
which together with Eq.~(\ref{T0b}) at $\lambda^{2} \ll 1$ yields
\begin{equation}
    F_{\sigma}\! \left( \frac{\pi}{2} \right) = -\sigma \rho
    \frac{e^{a}[1 + F_{\sigma}(\pi(1-\sigma)/2)] \tilde{h}^{2}}
    {2[g(\tilde{ \omega}) + \sigma \rho a\tilde{h}^{2}]},
    \label{F2b}
\end{equation}
where $g(\tilde{ \omega}) = \tilde{\omega}(\lambda^{2} + \tilde{
\omega}^{2}) /(1 + 2\tilde{\omega}^{2})$ and $F_{\sigma} (\pi(1-
\sigma)/2)$ is defined in Eq.~(\ref{F1}).

\subsubsection{Frequency dependence of the lifetime}
\label{FrDep}

Now we are in a position to determine the lifetime (\ref{Pmod}) of
the P mode. Using Eq.~(\ref{appr}) and the fact that
$\mathcal{T}_{\sigma 0}$ does not depend on $\theta_{\sigma}'$ and
$\psi_{\sigma}'$, the desired lifetime can be expressed as
\begin{equation}
    \mathcal{T}_{\sigma} = \mathcal{T}_{\sigma 0} +
    2 \big(\mathcal{T}_{\sigma 1}^{+} \cos \Psi_{\sigma}[\rho] -
    \mathcal{T}_{\sigma 1}^{-} \sin \Psi_{\sigma}[\rho]\big)
    \big|_{\theta_{\sigma}' = \Theta_{\sigma}[\rho]}.
    \label{LT1}
\end{equation}
If the field amplitude $\tilde{h}$ is small enough, then $\cos
\Theta_{ \sigma}[\rho]$ in Eq.~(\ref{st1}) can be replaced by
$\sigma$. One can easily check that in this case
\begin{equation}
    \tan \Psi_{\sigma}[\rho] = -\rho \frac{\lambda \tilde{
    \omega}} {1+\lambda^{2}- \sigma \rho \tilde{\omega}},
    \label{Psi}
\end{equation}
and the precession angle $\Theta_{\sigma} [\rho]$ is given by
\begin{equation}
    \Theta_{\sigma}[\rho] = \frac{\pi}{2}(1 - \sigma) +
    \sigma \tilde{h} \sqrt{ \frac{1+\lambda^{2}}{
    (1-\sigma \rho \tilde{\omega})^{2}+\lambda^{2}}}.
    \label{Theta}
\end{equation}
The last result shows that the functions $\mathcal{T}_{\sigma
1}^{\pm}$ in Eq.~(\ref{LT1}) must be taken from Eq.~(\ref{solT+-})
with $\eta_{\sigma} = \sigma \Theta_{\sigma}[\rho] + \pi(1-
\sigma)/2$. But, according to Eq.~(\ref{AsT0}), the term $\mathcal
{T}_{\sigma 0}$ contains an additional factor $e^{a}$, and so the
second term in Eq.~(\ref{LT1}) which describes the dependence of
$\mathcal {T}_{\sigma}$ on $\Psi_{\sigma}[\rho]$ can be safely
neglected at $a\gg 1$.

Thus, using Eqs.~(\ref{AsT0}), (\ref{F1}), and (\ref{F2b}), for the
lifetime of the P mode we obtain
\begin{equation}
    \mathcal {T}_{\sigma } = \frac{e^{a}}{\lambda}
    \sqrt{\frac{\pi}{a}}\,R_{\sigma\rho}(\tilde{\omega})
    S_{\sigma\rho}(\tilde{\omega}),
    \label{LT2}
\end{equation}
where
\begin{equation}
    R_{\sigma \rho}(\tilde{\omega}) = 1- \frac{a
    \tilde{h}^{2}} {(1-\sigma\rho \tilde{\omega})^{2}
    + \lambda^{2} + a\tilde{h}^{2}}
    \label{R}
\end{equation}
and
\begin{equation}
    S_{\sigma \rho}(\tilde{\omega}) = 1 -
    \frac{\sigma \rho a\tilde{h}^{2}} {g(\tilde{ \omega}) +
    \sigma \rho a\tilde{h}^{2}}.
    \label{S}
\end{equation}
If $\sigma \rho = +1$ then Eq.~(\ref{LT2}) correctly describes the
frequency dependence of the lifetime in the vicinity of the point
$\tilde{\omega} =1$ and at $\tilde{\omega} \gg 1$. Since
$a\tilde{h}^{2} \ll 1$, $\lambda^{2} \ll 1$ and $g(1) \approx 1/3 $,
in the former case we obtain $\mathcal {T}_{\sigma } =
(e^{a}/\lambda) \sqrt{\pi/a} R_{+1}(\tilde{\omega})$. To put it
differently, the rotating magnetic field whose direction of rotation
coincides with the direction of the natural precession decreases the
lifetime in a resonant manner, i.e., a resonant suppression of the
thermal stability of the P mode occurs. Taking into account that
$1-R_{+1}(\tilde{\omega}) \propto \tilde{ \omega}^{-2}$ and
$1-S_{+1}(\tilde{\omega}) \propto \tilde{ \omega}^{-1}$, in the
latter case Eq.~(\ref{LT2}) yields $\mathcal {T}_{\sigma } = (e^{a}/
\lambda) \sqrt{\pi/a} S_{+1}(\tilde{\omega})$. In contrast, at
$\sigma \rho = -1$ Eq.~(\ref{LT2}) describes only the high-frequency
behavior of the lifetime: $\mathcal {T}_{\sigma } = (e^{a}/ \lambda)
\sqrt{\pi/a} S_{-1}(\tilde{ \omega})$. Combining the last two
results, we obtain the expression
\begin{equation}
    \mathcal {T}_{\sigma} = \frac{e^{a}}{\lambda}
    \sqrt{\frac{\pi} {a}} \bigg(1-\sigma\rho
    \frac{2a\tilde{h}^{2}}{\tilde{\omega}}\bigg)
    \label{Thigh3}
\end{equation}
($\tilde{ \omega} \gg 1$), which shows that while at $\sigma \rho =
+1$ the rotating field suppresses the lifetime of the P mode, the
rotating field with $\sigma \rho = -1$ enhances it.

\subsubsection{Lifetime at zero frequency}
\label{Zero}

For the evaluation of the lifetime at $\tilde{\omega}=0$ it is
convenient to use the one-dimensional approximation, which consists
in replacing $\psi$ by $0$ in the magnetic energy (\ref{defW}). In
this case $u(\theta'_{\sigma}, \psi'_{\sigma}) = -(\lambda/2)\sin
2\theta'_{\sigma} + \lambda \tilde{h} \cos \theta'_{ \sigma}$, the
lifetime $\mathcal {T}_{\sigma}$ does not depend on $\psi'_{\sigma}$,
and the partial differential equation (\ref{MFPT1}) reduces to the
ordinary one:
\begin{equation}
    \frac{d^{2}\mathcal {T}_{\sigma}}{d\theta_{\sigma}'^{2}}
    + (\cot \theta_{\sigma}' -a\sin 2\theta_{\sigma}'
    + 2a\tilde{h}\cos \theta_{\sigma}') \frac{d\mathcal
    {T}_{\sigma}}{d\theta_{\sigma}'} = -\frac{2a}{\lambda}.
    \label{eq_Tzer}
\end{equation}
Its exact solution satisfying the conditions (\ref{absT}) and
(\ref{reflT}) [we note that for this equation the finiteness
condition (\ref{finT}) is equivalent to the reflecting boundary
condition (\ref{reflT})] can be written in the form
\begin{equation}
    \mathcal {T}_{\sigma} = \frac{2a}{\lambda} \int_{\cos
    \theta_{0}}^{\sigma \cos \theta_{\sigma}'} dx
    \frac{e^{-af(x)}}{1 - x^{2}} \int_{x}^{1}dy e^{af(y)},
    \label{Tzer1}
\end{equation}
where $f(x) = x^{2} + 2\tilde{h}\sqrt{1-x^{2}}$ ($|x|\leq 1$) is the
symmetric function with $\mathrm{min} f(x) = f(0) = 2\tilde{h}$,
$\mathrm{max}f(x) = f(\sqrt{1 - \tilde{h}^{2}}) = 1 + \tilde{h}^{2}$,
and $f(1)=1$.

As before, we are interested in the behavior of $\mathcal
{T}_{\sigma}$ at $a\gg 1$. In this case a small vicinity of the point
$x=0$ gives the main contribution to the first integral in
Eq.~(\ref{Tzer1}). This contribution depends not only on $a$ but also
on the parameter $a\tilde{h}$. In particular, if $a\tilde{h} \ll 1$
then, putting $x=0$ everywhere except for $e^{-af(x)}$, representing
$e^{-af(x)}$ in the vicinity of the point $x=0$ as $e^{-2a \tilde{h}
-a(1-\tilde{h})x^{2}} \approx (1 - 2a\tilde{h}) e^{-ax^{2}}$ and
extending the limits of integration to infinity, Eq.~(\ref{Tzer1})
yields
\begin{equation}
    \mathcal {T}_{\sigma} = \frac{2}{\lambda}\sqrt{\pi a}\,
    (1-2a\tilde{h}) \int_{0}^{1}dy e^{af(y)}.
    \label{Tzer2}
\end{equation}
Then, taking into account that
\begin{eqnarray}
    \int_{0}^{1}dy e^{af(y)} \!&\approx&\! e^{a}\int_{0}^{\infty}
    dz e^{-2az}(1 + 2^{3/2}a\tilde{h}\sqrt{z})
    \nonumber\\[6pt]
    \!&=&\! \frac{e^{a}
    }{2a}\,(1 + \sqrt{\pi a}\,\tilde{h})
    \label{relC1}
\end{eqnarray}
($a \gg 1$, $a\tilde{h} \ll 1$) and neglecting terms proportional to
$\sqrt{\pi a}\,\tilde{h}$, from Eqs.~(\ref{Tzer2}) and (\ref{relC1})
one finds
\begin{equation}
    \mathcal {T}_{\sigma} = \frac{e^{a}}{\lambda}
    \sqrt{\frac{\pi} {a}}\, (1-2a\tilde{h}).
    \label{Tzer3}
\end{equation}
Comparing this result with Eq.~(\ref{Thigh3}), we conclude that
$\mathcal {T}_{\sigma}|_{\tilde{\omega} = 0} < \mathcal
{T}_{\sigma}|_{\tilde {\omega} = \infty}$, and so at $\sigma \rho =
-1$ the frequency dependence of the lifetime has a maximum exceeding
the limiting value $\mathcal {T}_{\sigma}|_{\tilde {\omega} = \infty}
= (e^{a}/\lambda) \sqrt{\pi/a}$.

It should be noted that the lifetime $\mathcal {T}_{\sigma}|_{
\tilde{\omega} = 0}$ strongly decreases with increasing $\tilde{h}$.
For example, if $\tilde{h} \sim 1/\sqrt{a}$ then, instead of
Eq.~(\ref{Tzer2}), we obtain
\begin{equation}
    \mathcal {T}_{\sigma} = \frac{2}{\lambda}\sqrt{\frac{\pi a}
    {1-\tilde{h}}}\, e^{-2a\tilde{h}} \int_{0}^{1}dy e^{af(y)}.
    \label{Tzer4}
\end{equation}
To evaluate the integral in Eq.~(\ref{Tzer4}), it is convenient to
divide the interval of integration  $(0,1)$ into two parts, $(0,
\sqrt{1 - \tilde{h} ^{2}})$ and $(\sqrt{1 - \tilde{h} ^{2}}, 1)$, and
apply the Laplace method.\cite{Olver} By this way, the corresponding
integrals can easily be evaluated yielding
\begin{eqnarray}
    \int_{0}^{\sqrt{1 - \tilde{h} ^{2}}}dy e^{af(y)} \!&\approx&\!
    \frac{\tilde{h}}{\sqrt{a}}\,e^{a(1 + \tilde{h}^{2})}
    \int_{0}^{\sqrt{a}/\tilde{h}}dz e^{-z^{2}}
    \nonumber\\[6pt]
    \!&=&\! \frac{\tilde{h}}{2} \sqrt{\frac{\pi}
    {a}}\, e^{a(1 + \tilde{h} ^{2})} \mathrm{erf}
    \bigg( \frac{\sqrt{a}}{\tilde{h}}\bigg)
    \label{relC2}
\end{eqnarray}
and
\begin{eqnarray}
    \int_{\sqrt{1 - \tilde{h} ^{2}}}^{1}dy e^{af(y)} \!
    &\approx&\! \frac{\tilde{h}}{\sqrt{a}}\,e^{a(1 +
    \tilde{h}^{2})} \int_{0}^{\sqrt{a}\tilde{h}/2}dz
    e^{-z^{2}}
    \nonumber\\[6pt]
    \!&=&\! \frac{\tilde{h}}{2} \sqrt{\frac{\pi}
    {a}}\, e^{a(1 + \tilde{h} ^{2})} \mathrm{erf}
    \bigg( \frac{\sqrt{a}\tilde{h}}{2}\bigg), \qquad
    \label{relC3}
\end{eqnarray}
where $\mathrm{erf}(z) = (2/\sqrt{\pi}) \int_{0}^{z} dx e^{-x^{2}}$
is the error function. Finally, using
Eqs.~(\ref{Tzer4})--(\ref{relC3}) and the approximate formula
$\mathrm{erf} (\sqrt{a}/ \tilde{h}) \approx 1$ ($\sqrt{a}/ \tilde{h}
\sim a \gg 1$), we find the following expression for the lifetime at
$\tilde{\omega} =0$:
\begin{equation}
    \mathcal {T}_{\sigma} = \frac{\pi \tilde{h}}{\lambda
    \sqrt{1 - \tilde{h}}} \, e^{a(1-\tilde{h})^{2}}
    \bigg[ 1 + \mathrm{erf} \bigg( \frac{\sqrt{a}\tilde{h}}{2}
    \bigg) \bigg],
    \label{Tzer5}
\end{equation}
which is valid if $a \gg 1$ and $\tilde{h}$ is of the order of
$1/\sqrt{a}$. It is not difficult to see that for these conditions
the strong inequality $\mathcal {T}_{\sigma} |_{\tilde {\omega} = 0}
\ll \mathcal {T}_{\sigma} |_{\tilde {\omega} = \infty}$ holds. Taking
into account also Eq.~(\ref{Thigh3}), we can conclude that $\mathcal
{T}_{\sigma}$ as a function of $\tilde{\omega}$ at $\sigma \rho =-1$
has a local maximum (see Sec.~\ref{Simulat}).

\section{NUMERICAL RESULTS}
\label{NumRes}
\subsection{Precessional modes of the magnetic moment}
\label{Prec}

The analytical results suggest that the numerical analysis of the
lifetimes of the precessional modes should start with the study of
these modes without thermal fluctuations. More precisely, it is
necessary (i) to determine the conditions when for a given rotating
field one precessional mode exists in the up state of the magnetic
moment and one in the down state, and (ii) to study the steady-state
properties of these modes. To solve these problems, we use the
Landau-Lifshitz equation written in the form of Eq.~(\ref{L-L2}) with
$a=\infty$. In general, the finite state of the magnetic moment
(i.e., state at $\tilde{t} \to \infty$) depends not only on the
parameters $\tilde{h}$, $\tilde{\omega}$ and $\rho$ characterizing
the rotating field, but also on how this field is switched on. In
particular, a sharp switching of the rotating field induces dynamical
effects that may result in a change of the initial state
$\sigma$.\cite{LPRB} Although these effects can be important for
applications, they are out of our scope here. Therefore, to minimize
the role of dynamical effects, we assume that the switch-on of the
rotating field is slow enough.

In order to determine the character of the precessional modes for a
given rotating field, the following numerical scheme is used. First,
the field amplitude is discretized as $\tilde{h} = n \Delta
\tilde{h}$, where $n=1,2,\ldots$ and $\Delta \tilde{h}$ is the
amplitude increment. Then, putting $n=1$ and using the initial
conditions $\theta(0) = \pi(1- \sigma)/2 + \sigma 10^{-4}$ and
$\psi(0) = 0$, the fourth order Runge-Kutta method with a time step
$\Delta\tilde{t} = 10^{-3}$ is applied to solve Eq.~(\ref{L-L2}) at
$a = \infty$ on the time interval $[0, \tilde{t}_{\rm{m}}]$
($\tilde{t}_{ \rm{m}} \gg \tilde{t}_{ \rm{rel}} = 2/\lambda$). It is
assumed that at $\tilde{t} = \tilde{t}_{ \rm{m}}$ the field amplitude
jumps from $\Delta \tilde{h}$ to $2\Delta \tilde{h}$, i.e., $n$
becomes equal to two, and Eq.~(\ref{L-L2}) is solved again on the
interval $[0, \tilde{t}_{ \rm{m}}]$. But now the initial conditions
are the solutions obtained for $\tilde{t} = \tilde{t}_{\rm{m}}$ at
the previous stage: $\theta(0) |_{n=2} = \theta( \tilde{t}_{ \rm{m}})
|_{n=1}$ and $\psi(0)|_{n=2} = \psi(\tilde{t}_{ \rm{m}}) |_{n=1}$.
Continuing this procedure, Eq.~(\ref{L-L2}) can be solved for an
arbitrary $n$. It is worth to note that since $\tilde{t}_{\rm{m}} \gg
\tilde{t}_{ \rm{rel}}$ the solutions of this equation at $\tilde{t}
\sim \tilde{t}_{\rm{m}}$ are expected to be quite close to the
steady-state solutions $\Theta_{\sigma} (\tilde{t})$ and
$\Psi_{\sigma} (\tilde{t})$.

\begin{figure}
    \centering
    \includegraphics[totalheight=5cm]{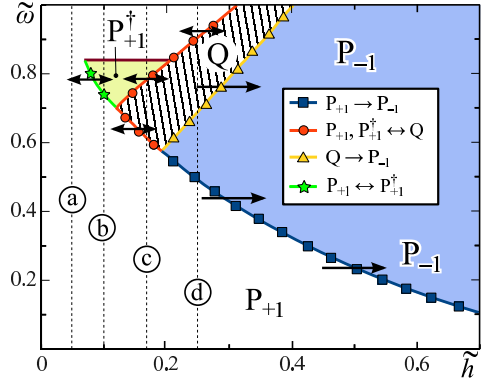}
    \caption{\label{fig1} (Color online) Diagram of the
    precessional modes for $\sigma \rho = +1$.
    The regions in the $\tilde{h}$-$\tilde{\omega}$ plane where
    different P modes exist at $\sigma \rho = +1$ are denoted as
    $\rm{P}_{+1}$ (white) and $\rm{P}_{+1}^{\dag}$ (light-green).
    The Q mode is realized in the white shaded region.
    In the region denoted as $\rm{P}_{-1}$ (blue) the stable
    precessional modes with $\sigma \rho = +1$ do not exist.
    Here, only the P mode with $\sigma \rho = -1$ is
    realized. The vertical dotted lines (a), (b), (c) and
    (d) correspond to $\tilde{h} = 0.05, \; 0.1, \; 0.18$
    and $0.25$, respectively. }
\end{figure}

Using the above procedure with $\Delta \tilde{h} = 10^{-2}$, $\tilde{
t}_{\rm{m}} = 10^{3}$ and $\lambda = 0.15$ [this value of $\lambda$,
which belongs to the interval $(0.01, 0.22)$ of typical values of the
damping parameter in the case of Co samples,\cite{BAW} is used in all
our numerical calculations], we determined the character of the
precessional modes for a wide range of parameters characterizing the
rotating field. It is established that if $\sigma \rho = -1$ then
only the P mode is realized for all $\tilde{h}$ and $\tilde{\omega}$.
In contrast, the precessional modes at $\sigma \rho = +1$ exhibit a
much more complex behavior. The results related to the character of
these modes are summarized in the diagram shown in Fig.~\ref{fig1}.
We note that the difference between two P modes, which exist in the
regions $\rm{P}_{+1}$ and $\rm{P}_{+1}^{\dag}$, is that the
precession angle $\Theta_{\sigma}$ as a function of $\tilde{h}$ is
discontinuous at the boundary between them. It should also be
emphasized that the transitions between the modes with $\sigma \rho =
+1$, which occur under changing the field amplitude $\tilde{h}$, are
reversible. For clearness of presentation, this fact is illustrated
by the horizontal bidirectional arrows. In contrast, the transitions
to the P mode with $\sigma \rho = -1$ are irreversible (they are
depicted by the horizontal unidirectional arrows).
Moreover, using Eq.~(\ref{st1}) and the stability criterion for the P
mode,\cite{DLHT} we independently confirmed the correctness of this
diagram by calculating the lines that separate the regions with
$\sigma \rho = +1$.

Since the rotating field is switched on during the time interval of
duration $\tilde{t} _{0} = (n-1)10^{3}$, the above procedure is time
consuming. However, in view of the importance of the diagram of the
precessional modes for the problem of lifetimes, its use for the
precise determination of this diagram is quite acceptable. At the
same time, the application of this method to the study of the
steady-state properties of a given mode, whose character is already
known from the diagram, is clearly redundant. Therefore, to reduce
the computational time, next we use a modified numerical procedure
with $\tilde{ t}_{\rm{m}} = \Delta \tilde{t}$ and $\Delta \tilde{h} =
\tilde{h} \Delta \tilde{t}/50$ leading to $\tilde{t}_{0} = \Delta
\tilde{t}\, \tilde{h}/\Delta \tilde{h} =50$.

\begin{figure}
    \centering
    \includegraphics[totalheight=5cm]{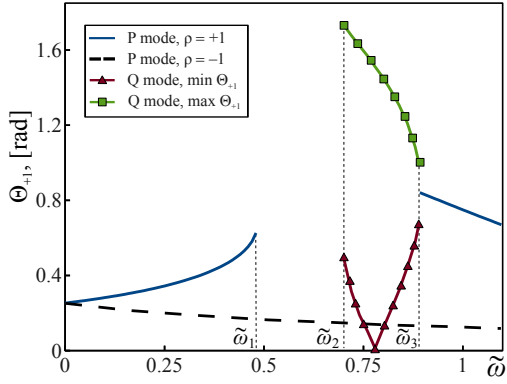}
    \caption{\label{fig2} (Color online) Frequency dependence
    of the precession angle $\Theta_{+1}$ for different modes
    that exist at $\tilde{h} = 0.25$. The frequencies $\tilde{
    \omega}_{1} = 0.49$, $\tilde{\omega}_{2} = 0.70$ and
    $\tilde{\omega}_{3} = 0.89$ are the coordinates of the
    points in which the vertical dotted line (d), see
    Fig.~\ref{fig1}, crosses the boundaries of the diagram
    ($\tilde{\omega}_{1}$, $\tilde{\omega}_{2}$, and
    $\tilde{\omega}_{3}$ depend on $\tilde{h}$).
    The green line (with squares) and brown line (with
    triangles) show the frequency dependence of $\max{
    \Theta_{+1}(\tilde{t})}$ and $\min{\Theta_{+1}
    (\tilde{t})}$, respectively, in the case of Q mode.}
\end{figure}

\begin{figure}
    \centering
    \includegraphics[totalheight=5cm]{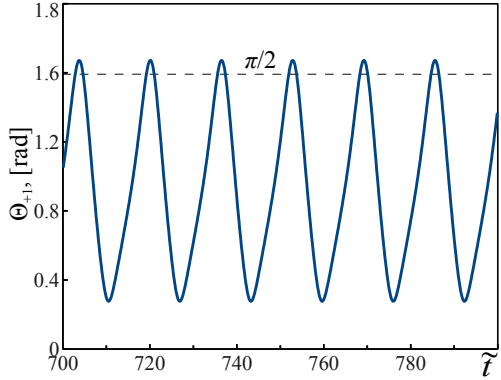}
    \caption{\label{fig3} (Color online) Time dependence
    of the precession angle $\Theta_{+1}(\tilde{t})$ in the
    case of Q mode. The parameters of the rotating field are
    as follows: $\rho = +1$, $\tilde{\omega} = 0.725$, and
    $\tilde{h} = 0.25$.}
\end{figure}

\begin{figure}
    \centering
    \includegraphics[totalheight=5cm]{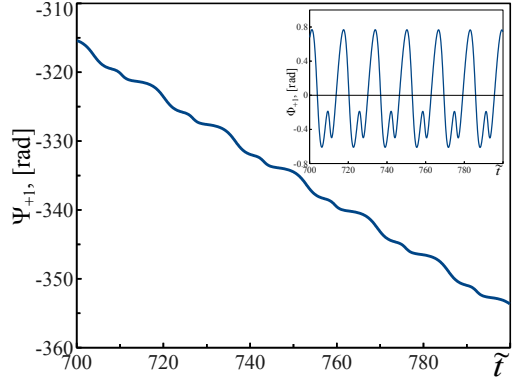}
    \caption{\label{fig4} (Color online) Time dependence of
    the difference of phases $\Psi_{+1} (\tilde{t}) = -\nu
    \tilde{t} + \Phi_{+1} (\tilde{t})$ in the case of Q mode.
    Insert: time dependence of the function $\Phi_{+1}
    (\tilde{t})$. The parameters of the rotating field are
    the same as in Fig.~\ref{fig3} and $\nu = 0.38$.}
\end{figure}

\begin{figure}
    \centering
    \includegraphics[totalheight=5cm]{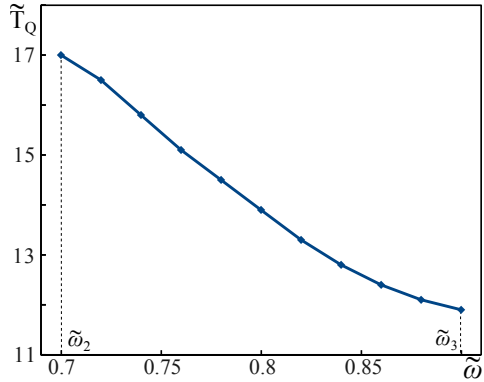}
    \caption{\label{fig5} (Color online) Frequency dependence
    of the period $\tilde{T}_{Q}$ of the precession angle
    $\Theta_{+1} (\tilde{t})$ in the case of Q mode.
    The rotating field parameters are taken as $\rho =+1$
    and $\tilde{h} = 0.25$.}
\end{figure}

Figure \ref{fig2} shows the frequency dependence of the precession
angle $\Theta_{+1}$  for different precessional modes.
The dashed (black) and solid (blue) lines correspond to the P modes
with $\sigma \rho =-1$ and $\sigma \rho =+1$, respectively. In
accordance with Fig.~\ref{fig1}, the Q mode occurs at $\tilde{
\omega}_{2} < \tilde{ \omega} < \tilde{ \omega}_{3}$ (we recall that
this mode can exist only if $\sigma \rho =+1$). For this mode, the
time dependence of the precession angle $\Theta_{+1} (\tilde{t})$ and
the time dependence of the difference of phases $\Psi_{+1}
(\tilde{t}) = -\nu\tilde{t} + \Phi_{+1} (\tilde{t})$ are illustrated
in Figs.~\ref{fig3} and \ref{fig4}.
We note that for a given set of parameters $\max{ \Theta_{+1}
(\tilde{t} )}>\pi/2$, i.e., the precession angle $\Theta_{\sigma}
(\tilde{t})$ in the case of Q mode can cross the anisotropy barrier.
Moreover, since $\sin \Theta_{\sigma} (\tilde{t})$ and $\sin
\Psi_{\sigma} (\tilde{t})$ have the same period $\tilde{T}_{Q}$, the
parameter $\nu$ and the period $\tilde{T}_{Q}$ (its frequency
dependence illustrates Figs.~\ref{fig5}) are connected by the
condition $\nu = 2k\pi/ \tilde{T}_{Q}$, where $k$ is a nonnegative
integer
that depends on $\tilde{h}$ and $\tilde{ \omega}$. In particular, if
$\tilde{h} = 0.25$ then $k=1$ at $\tilde{ \omega}_{2} <\tilde{
\omega}< \tilde{ \omega}_{2}'$, where $\tilde{ \omega}_{2}' = 0.78$
is the solution of the equation $\min{\Theta_{+1}( \tilde{t})}
|_{\tilde{ \omega} = \tilde{ \omega}_{2}'} =0$ (see Fig.~\ref{fig2}),
and $k=0$ at $\tilde{ \omega}_{2}' <\tilde{ \omega}< \tilde{
\omega}_{3}$. It should also be stressed that in the last case the
function $\Phi_{+1} (\tilde{t})$ shows a cosine-like behavior, in
contrast to that shown in Fig.~\ref{fig4}.

\subsection{Simulated lifetimes and their properties}
\label{Simulat}

\begin{figure}
    \centering
    \includegraphics[totalheight=4.7cm]{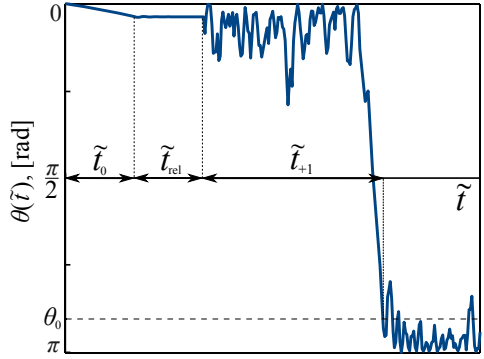}
    \caption{\label{fig6} (Color online) Schematic time
    dependence of the polar angle $\theta (\tilde{t})$
    in the regions $\rm{P}_{+1}$ and $\rm{P}_{+1}^{\dag}$
    shown in Fig.~\ref{fig1}. The change of the
    magnetic moment state $\sigma$ from $+1$ to $-1$
    occurs at $\tilde{t} = \tilde{t}_{0} + \tilde{t
    }_{\rm{rel}} + \tilde{t}_{+1}$, when $\theta
    (\tilde{t})$ reaches the angle $\theta_{0} =
    0.8\pi$ (the horizontal dashed line)
    for the first time. For a given trajectory
    $\theta (\tilde{t})$, the lifetime of the P mode
    in the state $\sigma= +1$ is equal to $\tilde{t
    }_{+1}$. Running $N\gg 1$ trajectories, the mean
    lifetime can be evaluated as $\mathcal {T}_{+1}
    = (1/N) \sum_{i=1}^{N} \tilde{t}_{+1}^{(i)}$.}
\end{figure}

As it was mentioned earlier, the thermal fluctuations can cause the
transitions between different precessional modes induced by a given
rotating field. According to the diagram in Fig.~\ref{fig1}, these
modes, one in the up state of the magnetic moment and the other in
the down state, exist only if the amplitude and frequency of the
rotating field belong to the region $P_{+1}$, $P_{+1}^{\dag}$ or Q.
In this case, the lifetime of a given mode can be calculated from the
numerical solution of the stochastic equations (\ref{L-L2}). Our
testing calculations showed that the solution of these equations by
the Euler method gives practically the same lifetime obtained by the
fourth order Runge-Kutta method. But in the first case the
calculation time is almost four times less. Therefore, because the
procedure of determining the frequency dependence of the lifetime is
extremely time-consuming, we used the Euler method. The time step
$\Delta \tilde{t}$ is chosen to be $10^{-3}$ and the initial
conditions are given by $\theta_{ \sigma}(0) = \pi(1- \sigma)/2 +
\sigma 10^{-2}$ and $\psi_{ \sigma}(0) = 0$. To prevent the
appearance of singularities in Eq.~(\ref{L-L2}) at $\theta_{\sigma} =
\pi(1- \sigma)/2$, we assume that the point $\theta_{\sigma} = \pi(1-
\sigma)/2 + \sigma 10^{-3}$ acts on the process $\theta_{\sigma}
(\tilde{t})$ as a reflecting screen. Since our interest here is the
lifetimes of the precessional modes \textit{reaching} the steady
state, the thermal fluctuations are switched on at $\tilde{t} =
\tilde{t}_{\rm{st}}$ (see Sec.~\ref{Def}) with $\tilde{t}_{ \rm{st}}
= \tilde{t}_{0} + \tilde{t}_{\rm{rel}}$ (see Fig.~\ref{fig6}). In the
case of P modes we chose $\tilde{t}_{ \rm{st}} = 10^{2}$, while for
the Q mode $\tilde{t}_{ \rm{st}} \in [10^{2}, 10^{2} +
\tilde{T}_{Q}]$. In the latter case, the thermal fluctuations can be
switched on at a certain instant of time, e.g., when the precession
angle $\Theta_{ \sigma} (\tilde{t})$ reaches maximum or minimum.
Finally, in all our numerical simulations the parameter
$a=mH_{a}/2k_{B}T$ is chosen to be ten.

\begin{figure}
    \centering
    \includegraphics[totalheight=5cm]{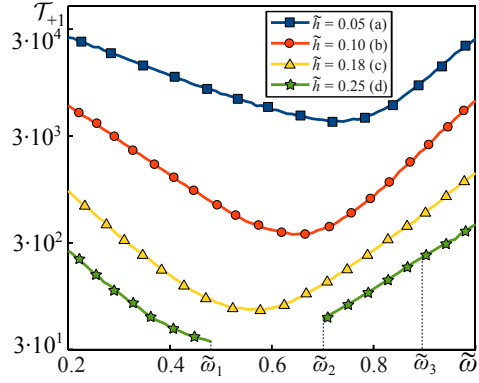}
    \caption{\label{fig7} (Color online) Frequency dependencies
    of the lifetime of the precessional modes induced by the
    rotating field with $\rho = +1$ in the up state
    ($\sigma = +1$) of the magnetic moment.}
\end{figure}

\begin{figure}
    \centering
    \includegraphics[totalheight=5cm]{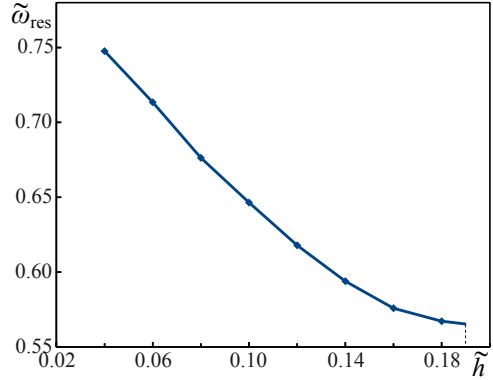}
    \caption{\label{fig8} (Color online) Dependence of the
    resonant frequency of the lifetime $\mathcal {T}_{+1}$
    on the rotating field amplitude.}
\end{figure}

In Fig.~\ref{fig7}, we show the frequency dependencies of the
lifetime $\mathcal {T}_{+1}$ for the rotating field amplitudes
indicated in Fig.~\ref{fig1}. Each point of these curves is
determined by running $N = 10^{4}$ trajectories of the polar angle
$\theta (\tilde{t})$ (see the caption to Fig.~\ref{fig6}).
If $\tilde{h} <0.19$ then the dependence of $\mathcal {T}_{+1}$ on
$\tilde{\omega}$ is continuous and exhibits a resonant minimum at
$\tilde{ \omega} = \tilde{ \omega}_{ \rm{res}}$. The plot of the
resonant frequency $\tilde{ \omega}_{ \rm{res}}$ versus the field
amplitude $\tilde{h}$ is shown in Fig.~\ref{fig8}. In contrast, if
$\tilde{h}>0.19$ then $\mathcal {T}_{+1}$ is discontinuous: at
$\tilde{\omega} \in (\tilde{ \omega}_{1}, \tilde{\omega}_{2})$ the
function $\mathcal {T}_{+1}$ does not exist. This result is a
consequence of the fact that in the region $\rm{P}_{-1}$ (see
Fig.~\ref{fig1}) there are no stable precessional modes with $\sigma
\rho = +1$.

\begin{figure}
    \centering
    \includegraphics[totalheight=5cm]{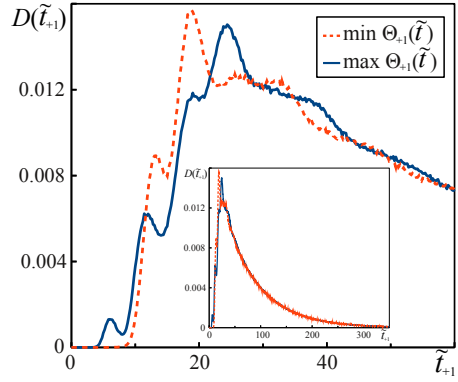}
    \caption{\label{fig9} (Color online) Probability density
    functions of the lifetime of the Q mode for two values of
    $\tilde{t}_{\rm{st}}$. The blue solid and red dashed
    lines correspond to such  $\tilde{t}_{\rm{st}}$ that
    $\Theta_{+1} (\tilde{t}_{\rm{st}}) = \max \Theta_{+1}
    (\tilde{t})$ and  $\Theta_{+1} (\tilde{t}_{\rm{st}}) =
    \min \Theta_{+1} (\tilde {t})$, respectively. Insert:
    the same density functions in a larger time scale.
    The parameters of the rotating field are as
    follows: $\rho = +1$, $\tilde{\omega} = 0.75$, and
    $\tilde{h} = 0.25$. }
\end{figure}

One more important feature of the lifetime $\mathcal {T}_{+1}$ is
that it is practically not sensitive to changing the character of the
precessional modes. In particular, for $\tilde{h} =0.1$ the frequency
dependence of $\mathcal {T}_{+1}$ (see the red line with circles in
Fig.~\ref{fig7}) is continuous at the point $\tilde{ \omega} = 0.74$
(i.e., point separating the regions $\rm{P}_{+1}$ and $\rm{P}_{+1
}^{\dag}$ at $\tilde{h} =0.1$), while the precession angle $\Theta
_{+1}$ is discontinuous. This insensibility of the lifetime to
changing the precessional modes with changing the field frequency
$\tilde{ \omega}$ is especially surprising when the Q mode appears.
For example, even at $\tilde{h} = 0.25$, when the precession angle
can cross the anisotropy barrier (see Fig.~\ref{fig3}), the character
of the frequency dependence of $\mathcal {T}_{+1}$ (see the green
line with stars in Fig.~\ref{fig7}) is changed at $\tilde{ \omega} =
\tilde{ \omega}_{3}$ so small that it is not visible on this scale.
Moreover, the lifetime of the Q mode does almost not depend on
$\tilde{t}_{ \rm{st}} \in [10^{2}, 10^{2} + \tilde{T}_{Q}]$. This
result is counterintuitive because different $\tilde{t}_{ \rm{st}}$
may correspond to very different values of the precession angle
$\Theta_{+1} (\tilde{t}_{\rm{st}})$. Therefore, in order to get more
insight into the problem, we calculated the probability density
function $D(\tilde{t}_{+1})$ of the lifetime of the Q mode for two
special values of $\tilde{t}_{\rm{st}}$ which correspond to $\min
\Theta_{+1} (\tilde{t})$ and $\max \Theta_{+1} (\tilde{t})$,
respectively. As seen from Fig.~\ref{fig9}, these density functions
are somewhat different from each other only if $\tilde{t}_{+1}
\lesssim \tilde{t}_{\rm{ rel}}$. In this region the probability
density $D(\tilde{t }_{+1})$ depends on $\tilde{t}_{ \rm{st}}$ and
exhibits local minima and maxima which come from a complex behavior
of the Q mode. In contrast, at $\tilde{t}_{+1} \gg \tilde{t}_{\rm{
rel}}$ the memory about the chosen value of $\tilde{t}_{ \rm{st}}$
and periodicity of $\Theta_{+1} (\tilde{t})$ and $\Phi_{+1}
(\tilde{t})$ is lost. As a consequence, in this region the difference
between density functions vanishes and the local minima and maxima
disappear. Such behavior of $D(\tilde{t}_{+1})$ on $\tilde{t}_{
\rm{st}}$ confirms that the lifetime $\mathcal {T}_{+1} = \int_{0}^{
\infty} d \tilde{t }_{+1}\, \tilde{t}_{+1} D(\tilde{t }_{+1})$
practically does not depend on $\tilde{t}_{ \rm{st}}$.

\begin{figure}
    \centering
    \includegraphics[totalheight=5cm]{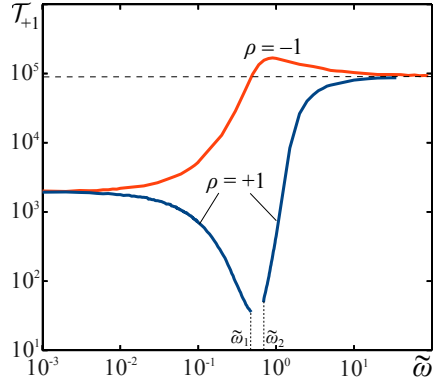}
    \caption{\label{fig10} (Color online) Frequency dependencies
    of the lifetime of the precessional modes in the up state
    ($\sigma = +1$) of the magnetic moment. It is assumed
    that $\tilde{h} = 0.25$ for both clockwise
    ($\rho = -1$) and counterclockwise ($\rho = +1$)
    rotation of the magnetic field. }
\end{figure}

Finally, the influence of the direction of field rotation on the
frequency dependence of the lifetime of the precessional modes is
illustrated in Fig.~\ref{fig10}. In accordance with our analytical
results, the rotating field with $\rho = +1$ and $\rho = -1$
influences the lifetime $\mathcal {T}_{+1}$ in a different way.
Specifically, $\mathcal {T}_{+1}$ as a function of $\tilde{ \omega}$
at $\rho = +1$ displays a deep minimum, while at $\rho = -1$ it shows
a pronounced maximum.
This difference in the behavior of the lifetime results from that the
rotating fields with different $\rho$ act on the magnetic moment in a
given state $\sigma$ quite differently. From a physical point of
view, the reason is that the magnetic moment has a definite direction
of the natural precession. We note also that the numerical data
$\mathcal {T}_{+1}|_{\tilde{\omega} \to \infty} = 9.1\times 10^{4}$
and $\mathcal {T}_{+1}|_{\tilde{\omega} \to 0} = 1.97\times 10^{3}$
are in a good agreement with the analytical results $8.2\times
10^{4}$ and $2.38\times 10^{3}$ obtained from the asymptotic formulas
(\ref{Thigh3}) and (\ref{Tzer5}), respectively. Some difference
between them can be caused by that the asymptotic formulas, which
were obtained at $a \to \infty$, are applied to $a=10$.

\section{CONCLUSIONS}
\label{Con}

We have studied in detail the thermal stability of the precessional
modes of the nanoparticle magnetic moment induced by the rotating
magnetic field whose plane of rotation is perpendicular to the easy
axis of the nanoparticle. If the direction of field rotation and the
direction of the natural precession of the magnetic moment are
opposite, i.e., if the condition $\sigma \rho = -1$ holds, then only
periodic (P) stable mode is induced by this field. In contrast, if
the above mentioned directions coincide, i.e., if $\sigma \rho = +1$,
then the magnetic moment exhibits a much more complicated behavior.
The numerical solution of the deterministic Landau-Lifshitz equation
in the long-time limit has shown that, depending on the rotating
field amplitude and frequency, in this case the magnetic moment can
be in one of two P modes, in the quasi-periodic (Q) mode, or even be
unstable. These results obtained in the absence of thermal
fluctuations have been collected in the diagram shown in Fig. 1.

If the amplitude and frequency of the rotating field are chosen so
that a stable precessional mode exists in both up ($\sigma = +1$) and
down ($\sigma = -1$) states of the magnetic moment, then the thermal
fluctuations can cause transitions between these modes. One of the
most important parameters characterizing these transitions is the
lifetime of a given mode. Since it can be naturally associated with
the mean first-passage time for the magnetic moment, we have used the
Fokker-Planck formalism to define this quantity and calculate its
properties. In particular, we have determined the boundary conditions
and transformation properties of the lifetime and have developed an
analytical method for finding its frequency dependence in the case of
high anisotropy barrier and small amplitudes of the rotating field.
Using this method, it has been shown that the rotating field (a)
slightly decreases (if $\sigma \rho = +1$) or increases (if $\sigma
\rho = -1$) the lifetime of the P mode at large frequencies and (b)
strongly decreases it (if $\sigma \rho = +1$) in the vicinity of the
Larmor frequency. We have also established that at zero frequency the
lifetime is always less than in the limit of large frequencies.

These analytical findings for the lifetime of the P mode have been
confirmed by our numerical simulations of the stochastic
Landau-Lifshitz equation. Moreover, the numerical simulations of this
equation for not too small amplitudes of the rotating field permitted
us to solve the problem of the lifetime of the Q mode. Since in this
case the precession angle is a periodic function of time which can
cross the anisotropy barrier, the solution of this problem is of
particular interest. It has turned out that, although the
precessional angle depends on time, the lifetime of the Q mode
practically does not depend on this time, i.e., on the precession
angle. We have also verified this result by calculating the lifetime
from the first-passage time distributions that correspond to
different values of the precession angle.

\section*{ACKNOWLEDGMENTS}
We are grateful to the Institute of Applied Physics NAS of Ukraine
for granting us access to their computing facilities.

\end{document}